\documentclass[
 10pt,
 twocolumn,
 showpacs,
 amsmath,amssymb,
 aps,
 prd,
 superscriptaddress,
 floatfix
]{revtex4-1}

\usepackage{graphicx}% Include figure files
\usepackage[pdftex,colorlinks,urlcolor=teal,linkcolor=blue,citecolor=red,pdfstartview=FitH]{hyperref} 
\usepackage{dcolumn}% Align table columns on decimal point
\usepackage{tabularx}
\usepackage{bm}% bold math
\usepackage{xcolor}
\usepackage{xspace}
\usepackage{amsmath}

\newcommand{\GeVcc}{\ensuremath{\,\mathrm{GeV}/\mathrm{c}^2}\xspace}

\begin{document}

\title{Projected Sensitivity of the SuperCDMS SNOLAB experiment}

\affiliation{Division of Physics, Mathematics, \& Astronomy, California Institute of Technology, Pasadena, CA 91125, USA}
\affiliation{Department of Physics, Durham University, Durham, DH1 3LE UK}
\affiliation{Fermi National Accelerator Laboratory, Batavia, IL 60510, USA}
\affiliation{Lawrence Berkeley National Laboratory, Berkeley, CA 94720, USA}
\affiliation{Department of Physics, Massachusetts Institute of Technology, Cambridge, MA 02139, USA}
\affiliation{School of Physical Sciences, National Institute of Science Education and Research, HBNI, Jatni - 752050, India}
\affiliation{Department of Physics \& Astronomy, Northwestern University, Evanston, IL 60208-3112, USA}
\affiliation{Pacific Northwest National Laboratory, Richland, WA 99352, USA}
\affiliation{Department of Physics, Queen's University, Kingston ON, Canada K7L 3N6}
\affiliation{Department of Physics, Santa Clara University, Santa Clara, CA 95053, USA}
\affiliation{SLAC National Accelerator Laboratory/Kavli Institute for Particle Astrophysics and Cosmology, Menlo Park 94025, CA}
\affiliation{Department of Physics, South Dakota School of Mines and Technology, Rapid City, SD 57701, USA}
\affiliation{Department of Physics, Southern Methodist University, Dallas, TX 75275, USA}
\affiliation{Department of Physics, Stanford University, Stanford, CA 94305, USA}
\affiliation{Department of Physics, Syracuse University, Syracuse, NY 13244, USA}
\affiliation{Department of Physics and Astronomy, and the Mitchell Institute for Fundamental Physics and Astronomy, Texas A\&M University, College Station, TX 77843, USA}
\affiliation{Instituto de F\'{\i}sica Te\'orica UAM/CSIC, Universidad Aut\'onoma de Madrid, 28049 Madrid, Spain}
\affiliation{Department of Physics \& Astronomy, University of British Columbia, Vancouver BC, Canada V6T 1Z1}
\affiliation{Department of Physics, University of California, Berkeley, CA 94720, USA}
\affiliation{Departments of Physics and Electrical Engineering, University of Colorado Denver, Denver, CO 80217, USA}
\affiliation{Department of Physics, University of California, Santa Barbara, CA 93106, USA}
\affiliation{Department of Physics, University of Evansville, Evansville, IN 47722, USA}
\affiliation{Department of Physics, University of Florida, Gainesville, FL 32611, USA}
\affiliation{School of Physics \& Astronomy, University of Minnesota, Minneapolis, MN 55455, USA}
\affiliation{Department of Physics, University of South Dakota, Vermillion, SD 57069, USA}
\affiliation{Department of Physics, University of Toronto, Toronto ON, Canada M5S 1A7}
\author{R.~Agnese} \affiliation{Department of Physics, University of Florida, Gainesville, FL 32611, USA}
\author{A.J.~Anderson} \affiliation{Department of Physics, Massachusetts Institute of Technology, Cambridge, MA 02139, USA}
\author{T.~Aramaki} \affiliation{SLAC National Accelerator Laboratory/Kavli Institute for Particle Astrophysics and Cosmology, Menlo Park 94025, CA}
\author{I.~Arnquist} \affiliation{Pacific Northwest National Laboratory, Richland, WA 99352, USA}
\author{W.~Baker} \affiliation{Department of Physics and Astronomy, and the Mitchell Institute for Fundamental Physics and Astronomy, Texas A\&M University, College Station, TX 77843, USA}
\author{D.~Barker} \affiliation{School of Physics \& Astronomy, University of Minnesota, Minneapolis, MN 55455, USA}
\author{R.~Basu~Thakur} \affiliation{Fermi National Accelerator Laboratory, Batavia, IL 60510, USA}\affiliation{Department of Physics, University of Illinois at Urbana-Champaign, Urbana, IL 61801, USA}
\author{D.A.~Bauer} \affiliation{Fermi National Accelerator Laboratory, Batavia, IL 60510, USA}
\author{A.~Borgland} \affiliation{SLAC National Accelerator Laboratory/Kavli Institute for Particle Astrophysics and Cosmology, Menlo Park 94025, CA}
\author{M.A.~Bowles} \affiliation{Department of Physics, South Dakota School of Mines and Technology, Rapid City, SD 57701, USA}
\author{P.L.~Brink} \affiliation{SLAC National Accelerator Laboratory/Kavli Institute for Particle Astrophysics and Cosmology, Menlo Park 94025, CA}
\author{R.~Bunker} \affiliation{Pacific Northwest National Laboratory, Richland, WA 99352, USA}
\author{B.~Cabrera} \affiliation{Department of Physics, Stanford University, Stanford, CA 94305, USA}
\author{D.O.~Caldwell} \affiliation{Department of Physics, University of California, Santa Barbara, CA 93106, USA}
\author{R.~Calkins} \affiliation{Department of Physics, Southern Methodist University, Dallas, TX 75275, USA}
\author{C.~Cartaro} \affiliation{SLAC National Accelerator Laboratory/Kavli Institute for Particle Astrophysics and Cosmology, Menlo Park 94025, CA}
\author{D.G.~Cerde\~no} \affiliation{Department of Physics, Durham University, Durham, DH1 3LE UK}\affiliation{Instituto de F\'{\i}sica Te\'orica UAM/CSIC, Universidad Aut\'onoma de Madrid, 28049 Madrid, Spain}
\author{H.~Chagani} \affiliation{School of Physics \& Astronomy, University of Minnesota, Minneapolis, MN 55455, USA}
\author{Y.~Chen} \affiliation{Department of Physics, Syracuse University, Syracuse, NY 13244, USA}
\author{J.~Cooley} \affiliation{Department of Physics, Southern Methodist University, Dallas, TX 75275, USA}
\author{B.~Cornell} \affiliation{Division of Physics, Mathematics, \& Astronomy, California Institute of Technology, Pasadena, CA 91125, USA}
\author{P.~Cushman} \affiliation{School of Physics \& Astronomy, University of Minnesota, Minneapolis, MN 55455, USA}
\author{M.~Daal} \affiliation{Department of Physics, University of California, Berkeley, CA 94720, USA}
\author{P.C.F.~Di~Stefano} \affiliation{Department of Physics, Queen's University, Kingston ON, Canada K7L 3N6}
\author{T.~Doughty} \affiliation{Department of Physics, University of California, Berkeley, CA 94720, USA}
\author{L.~Esteban} \affiliation{Instituto de F\'{\i}sica Te\'orica UAM/CSIC, Universidad Aut\'onoma de Madrid, 28049 Madrid, Spain}
\author{S.~Fallows} \affiliation{School of Physics \& Astronomy, University of Minnesota, Minneapolis, MN 55455, USA}
\author{E.~Figueroa-Feliciano} \affiliation{Department of Physics \& Astronomy, Northwestern University, Evanston, IL 60208-3112, USA}
\author{M.~Fritts} \affiliation{School of Physics \& Astronomy, University of Minnesota, Minneapolis, MN 55455, USA}
\author{G.~Gerbier} \affiliation{Department of Physics, Queen's University, Kingston ON, Canada K7L 3N6}
\author{M.~Ghaith} \affiliation{Department of Physics, Queen's University, Kingston ON, Canada K7L 3N6}
\author{G.L.~Godfrey} \affiliation{SLAC National Accelerator Laboratory/Kavli Institute for Particle Astrophysics and Cosmology, Menlo Park 94025, CA}
\author{S.R.~Golwala} \affiliation{Division of Physics, Mathematics, \& Astronomy, California Institute of Technology, Pasadena, CA 91125, USA}
\author{J.~Hall} \affiliation{Pacific Northwest National Laboratory, Richland, WA 99352, USA}
\author{H.R.~Harris} \affiliation{Department of Physics and Astronomy, and the Mitchell Institute for Fundamental Physics and Astronomy, Texas A\&M University, College Station, TX 77843, USA}
\author{T.~Hofer} \affiliation{School of Physics \& Astronomy, University of Minnesota, Minneapolis, MN 55455, USA}
\author{D.~Holmgren} \affiliation{Fermi National Accelerator Laboratory, Batavia, IL 60510, USA}
\author{Z.~Hong} \affiliation{Department of Physics \& Astronomy, Northwestern University, Evanston, IL 60208-3112, USA}
\author{E.~Hoppe} \affiliation{Pacific Northwest National Laboratory, Richland, WA 99352, USA}
\author{L.~Hsu} \affiliation{Fermi National Accelerator Laboratory, Batavia, IL 60510, USA}
\author{M.E.~Huber} \affiliation{Departments of Physics and Electrical Engineering, University of Colorado Denver, Denver, CO 80217, USA}
\author{V.~Iyer} \affiliation{School of Physical Sciences, National Institute of Science Education and Research, HBNI, Jatni - 752050, India}
\author{D.~Jardin} \affiliation{Department of Physics, Southern Methodist University, Dallas, TX 75275, USA}
\author{A.~Jastram} \affiliation{Department of Physics and Astronomy, and the Mitchell Institute for Fundamental Physics and Astronomy, Texas A\&M University, College Station, TX 77843, USA}
\author{M.H.~Kelsey} \affiliation{SLAC National Accelerator Laboratory/Kavli Institute for Particle Astrophysics and Cosmology, Menlo Park 94025, CA}
\author{A.~Kennedy} \affiliation{School of Physics \& Astronomy, University of Minnesota, Minneapolis, MN 55455, USA}
\author{A.~Kubik} \affiliation{Department of Physics and Astronomy, and the Mitchell Institute for Fundamental Physics and Astronomy, Texas A\&M University, College Station, TX 77843, USA}
\author{N.A.~Kurinsky} \affiliation{SLAC National Accelerator Laboratory/Kavli Institute for Particle Astrophysics and Cosmology, Menlo Park 94025, CA}
\author{A.~Leder} \affiliation{Department of Physics, Massachusetts Institute of Technology, Cambridge, MA 02139, USA}
\author{B.~Loer} \affiliation{Pacific Northwest National Laboratory, Richland, WA 99352, USA}
\author{E.~Lopez~Asamar} \affiliation{Department of Physics, Durham University, Durham, DH1 3LE UK}
\author{P.~Lukens} \affiliation{Fermi National Accelerator Laboratory, Batavia, IL 60510, USA}
\author{R.~Mahapatra} \affiliation{Department of Physics and Astronomy, and the Mitchell Institute for Fundamental Physics and Astronomy, Texas A\&M University, College Station, TX 77843, USA}
\author{V.~Mandic} \affiliation{School of Physics \& Astronomy, University of Minnesota, Minneapolis, MN 55455, USA}
\author{N.~Mast} \affiliation{School of Physics \& Astronomy, University of Minnesota, Minneapolis, MN 55455, USA}
\author{N.~Mirabolfathi} \affiliation{Department of Physics and Astronomy, and the Mitchell Institute for Fundamental Physics and Astronomy, Texas A\&M University, College Station, TX 77843, USA}
\author{R.A.~Moffatt} \affiliation{Department of Physics, Stanford University, Stanford, CA 94305, USA}
\author{J.D.~Morales~Mendoza} \affiliation{Department of Physics and Astronomy, and the Mitchell Institute for Fundamental Physics and Astronomy, Texas A\&M University, College Station, TX 77843, USA}
\author{J.L.~Orrell} \affiliation{Pacific Northwest National Laboratory, Richland, WA 99352, USA}
\author{S.M.~Oser} \affiliation{Department of Physics \& Astronomy, University of British Columbia, Vancouver BC, Canada V6T 1Z1}
\author{K.~Page} \affiliation{Department of Physics, Queen's University, Kingston ON, Canada K7L 3N6}
\author{W.A.~Page} \affiliation{Department of Physics \& Astronomy, University of British Columbia, Vancouver BC, Canada V6T 1Z1}
\author{R.~Partridge} \affiliation{SLAC National Accelerator Laboratory/Kavli Institute for Particle Astrophysics and Cosmology, Menlo Park 94025, CA}
\author{M.~Pepin} \affiliation{School of Physics \& Astronomy, University of Minnesota, Minneapolis, MN 55455, USA}
\author{A.~Phipps} \affiliation{Department of Physics, University of California, Berkeley, CA 94720, USA}
\author{S.~Poudel} \affiliation{Department of Physics, University of South Dakota, Vermillion, SD 57069, USA}
\author{M.~Pyle} \affiliation{Department of Physics, University of California, Berkeley, CA 94720, USA}
\author{H.~Qiu} \affiliation{Department of Physics, Southern Methodist University, Dallas, TX 75275, USA}
\author{W.~Rau} \affiliation{Department of Physics, Queen's University, Kingston ON, Canada K7L 3N6}
\author{P.~Redl} \affiliation{Department of Physics, Stanford University, Stanford, CA 94305, USA}
\author{A.~Reisetter} \affiliation{Department of Physics, University of Evansville, Evansville, IN 47722, USA}
\author{A.~Roberts} \affiliation{Department of Physics, University of South Dakota, Vermillion, SD 57069, USA}
\author{A.E.~Robinson} \affiliation{Fermi National Accelerator Laboratory, Batavia, IL 60510, USA}
\author{H.E.~Rogers} \affiliation{School of Physics \& Astronomy, University of Minnesota, Minneapolis, MN 55455, USA}
\author{T.~Saab} \affiliation{Department of Physics, University of Florida, Gainesville, FL 32611, USA}
\author{B.~Sadoulet} \affiliation{Department of Physics, University of California, Berkeley, CA 94720, USA}\affiliation{Lawrence Berkeley National Laboratory, Berkeley, CA 94720, USA}
\author{J.~Sander} \affiliation{Department of Physics, University of South Dakota, Vermillion, SD 57069, USA}
\author{K.~Schneck} \affiliation{SLAC National Accelerator Laboratory/Kavli Institute for Particle Astrophysics and Cosmology, Menlo Park 94025, CA}
\author{R.W.~Schnee} \affiliation{Department of Physics, South Dakota School of Mines and Technology, Rapid City, SD 57701, USA}
\author{B.~Serfass} \affiliation{Department of Physics, University of California, Berkeley, CA 94720, USA}
\author{D.~Speller} \affiliation{Department of Physics, University of California, Berkeley, CA 94720, USA}
\author{M.~Stein} \affiliation{Department of Physics, Southern Methodist University, Dallas, TX 75275, USA}
\author{J.~Street} \affiliation{Department of Physics, South Dakota School of Mines and Technology, Rapid City, SD 57701, USA}
\author{H.A.~Tanaka} \affiliation{Department of Physics, University of Toronto, Toronto ON, Canada M5S 1A7}
\author{D.~Toback} \affiliation{Department of Physics and Astronomy, and the Mitchell Institute for Fundamental Physics and Astronomy, Texas A\&M University, College Station, TX 77843, USA}
\author{R.~Underwood} \affiliation{Department of Physics, Queen's University, Kingston ON, Canada K7L 3N6}
\author{A.N.~Villano} \affiliation{School of Physics \& Astronomy, University of Minnesota, Minneapolis, MN 55455, USA}
\author{B.~von~Krosigk} \affiliation{Department of Physics \& Astronomy, University of British Columbia, Vancouver BC, Canada V6T 1Z1}
\author{B.~Welliver} \affiliation{Department of Physics, University of Florida, Gainesville, FL 32611, USA}
\author{J.S.~Wilson} \affiliation{Department of Physics and Astronomy, and the Mitchell Institute for Fundamental Physics and Astronomy, Texas A\&M University, College Station, TX 77843, USA}
\author{D.H.~Wright} \affiliation{SLAC National Accelerator Laboratory/Kavli Institute for Particle Astrophysics and Cosmology, Menlo Park 94025, CA}
\author{S.~Yellin} \affiliation{Department of Physics, Stanford University, Stanford, CA 94305, USA}
\author{J.J.~Yen} \affiliation{Department of Physics, Stanford University, Stanford, CA 94305, USA}
\author{B.A.~Young} \affiliation{Department of Physics, Santa Clara University, Santa Clara, CA 95053, USA}
\author{X.~Zhang} \affiliation{Department of Physics, Queen's University, Kingston ON, Canada K7L 3N6}
\author{X.~Zhao} \affiliation{Department of Physics and Astronomy, and the Mitchell Institute for Fundamental Physics and Astronomy, Texas A\&M University, College Station, TX 77843, USA}

\collaboration{SuperCDMS Collaboration}\noaffiliation

\date{\today}% It is always \today, today, but any date may be explicitly specified

\begin{abstract}
SuperCDMS SNOLAB will be a next-generation experiment aimed at directly detecting low-mass ($\le 10\GeVcc$) particles that may constitute dark matter by using cryogenic detectors of two types (HV and iZIP) and two target materials (germanium and silicon). The experiment is being designed with an initial sensitivity to nuclear recoil cross sections $\sim 1 \times 10^{-43}\,\mathrm{cm}^2$ for a dark matter particle mass of 1\GeVcc, and with capacity to continue exploration to both smaller masses and better sensitivities. 
The phonon sensitivity of the HV detectors will be sufficient to detect nuclear recoils from sub-GeV dark matter.
A detailed calibration of the detector response to low energy recoils will be needed to optimize running conditions of the HV detectors and to interpret their data for dark matter searches. Low-activity shielding, and the depth of SNOLAB, will reduce most backgrounds, but cosmogenically produced $^{3}$H and naturally occurring $^{32}$Si will be present in the detectors at some level. Even if these backgrounds are $\times$10 higher than expected, the science reach of the HV detectors would be over three orders of magnitude beyond current results for a dark matter mass of $1\GeVcc$. The iZIP detectors are relatively insensitive to variations in detector response and backgrounds, and will provide better sensitivity for dark matter particle masses ($\ge 5\GeVcc$). The mix of detector types (HV and iZIP), and targets (germanium and silicon), planned for the experiment, as well as flexibility in how the detectors are operated, will allow us to maximize the low-mass reach, and understand the backgrounds that the experiment will encounter. Upgrades to the experiment, perhaps with a variety of ultra-low-background cryogenic detectors, will extend dark matter sensitivity down to the ``neutrino floor'', where coherent scatters of solar neutrinos become a limiting background.

\end{abstract}

 % PACS, the Physics and Astronomy Classification Scheme.

\maketitle

%%
%% INTRODUCTION
%%

\section{Introduction}
\label{sec:Intro}

Understanding the nature of dark matter is one of the most compelling problems in cosmology, and the solution may well come from particle physics. For the past two decades, experimental attempts to directly detect particles that may constitute dark matter have focused on Weakly Interacting Massive Particles (WIMPs), such as those suggested by the supersymmetric model of particle physics, with favored masses in the 10\GeVcc to 10\,TeV/c$^2$ range. However, evidence for supersymmetry has not been forthcoming from the LHC experiments. More recent theoretical models such as asymmetric dark matter~\cite{Kaplan:2009ag} and dark sectors~\cite{Alexander:2016arXiv} suggest that a new experimental focus on low-mass dark matter particles may be productive. 

SuperCDMS SNOLAB is a next-generation experiment being designed to search for such low-mass ($\lesssim~10\GeVcc$) dark matter particles. Projected sensitivities for the experiment indicate that a thorough search for dark matter particles in this mass range is possible. This paper describes how these projected sensitivities are calculated and provides a parametric study of the dependence of these sensitivities on the main design drivers.  

%%
%% CONFIGURATION
%%

\section{Experimental Configuration}
\label{sec:Experiment}

The planned SuperCDMS SNOLAB experiment will be located approximately 2\,km underground within SNOLAB in Sudbury, Ontario, Canada. The SNOLAB rock overburden provides shielding against cosmic-ray secondaries equivalent to 6010 meters of water. The experiment will be located within the ``ladder lab'' drift at SNOLAB~\cite{Duncan:2010zz}.

\subsection{The SuperCDMS Detectors}
\label{sec:Detectors}
\begin{figure}
HV: \raisebox{-0.5\height}{\includegraphics[width=.65\columnwidth, trim={600 200 600 350},clip]{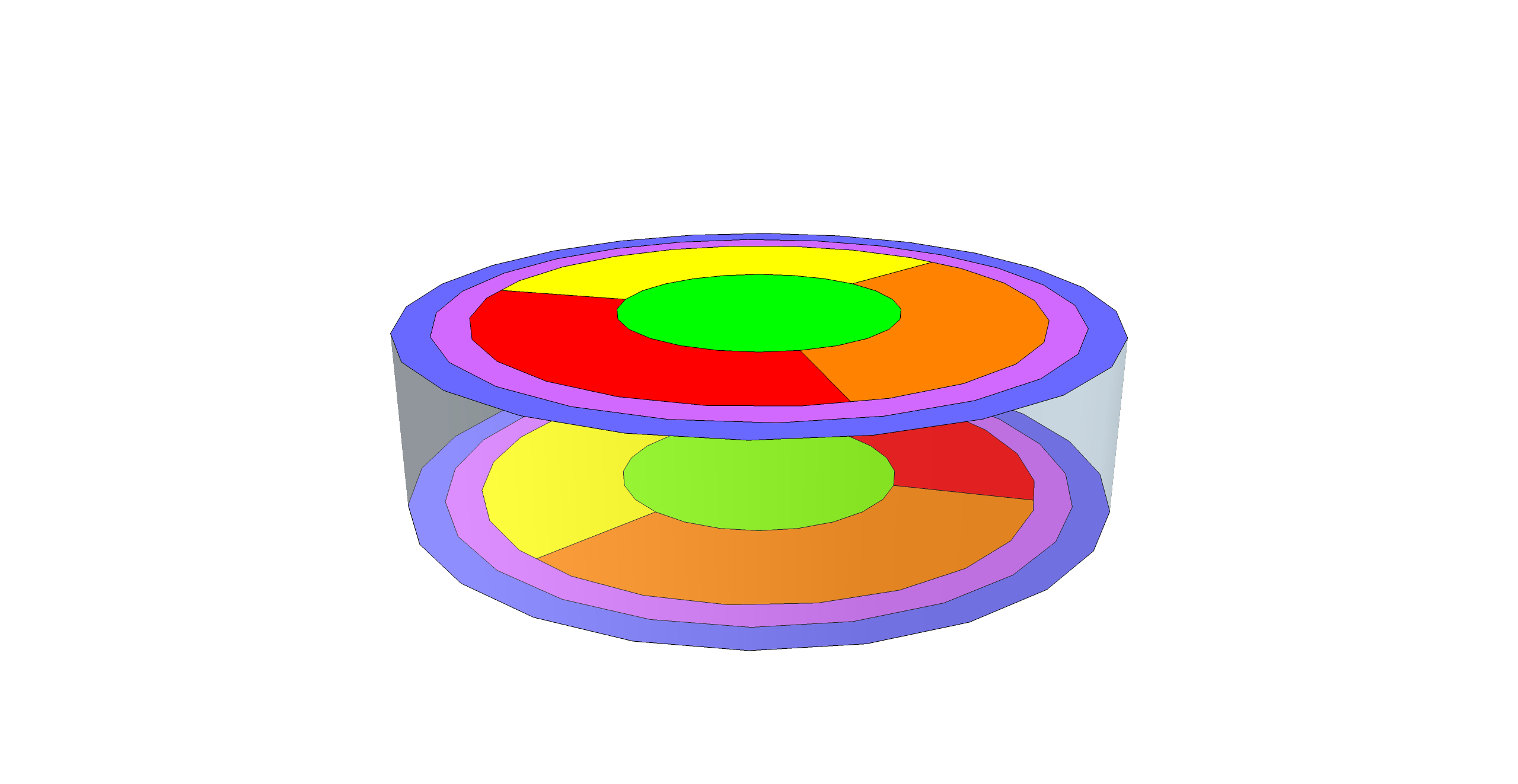}}\\
iZIP: \raisebox{-0.5\height}{\includegraphics[width=.65\columnwidth, trim={600 200 600 350},clip]{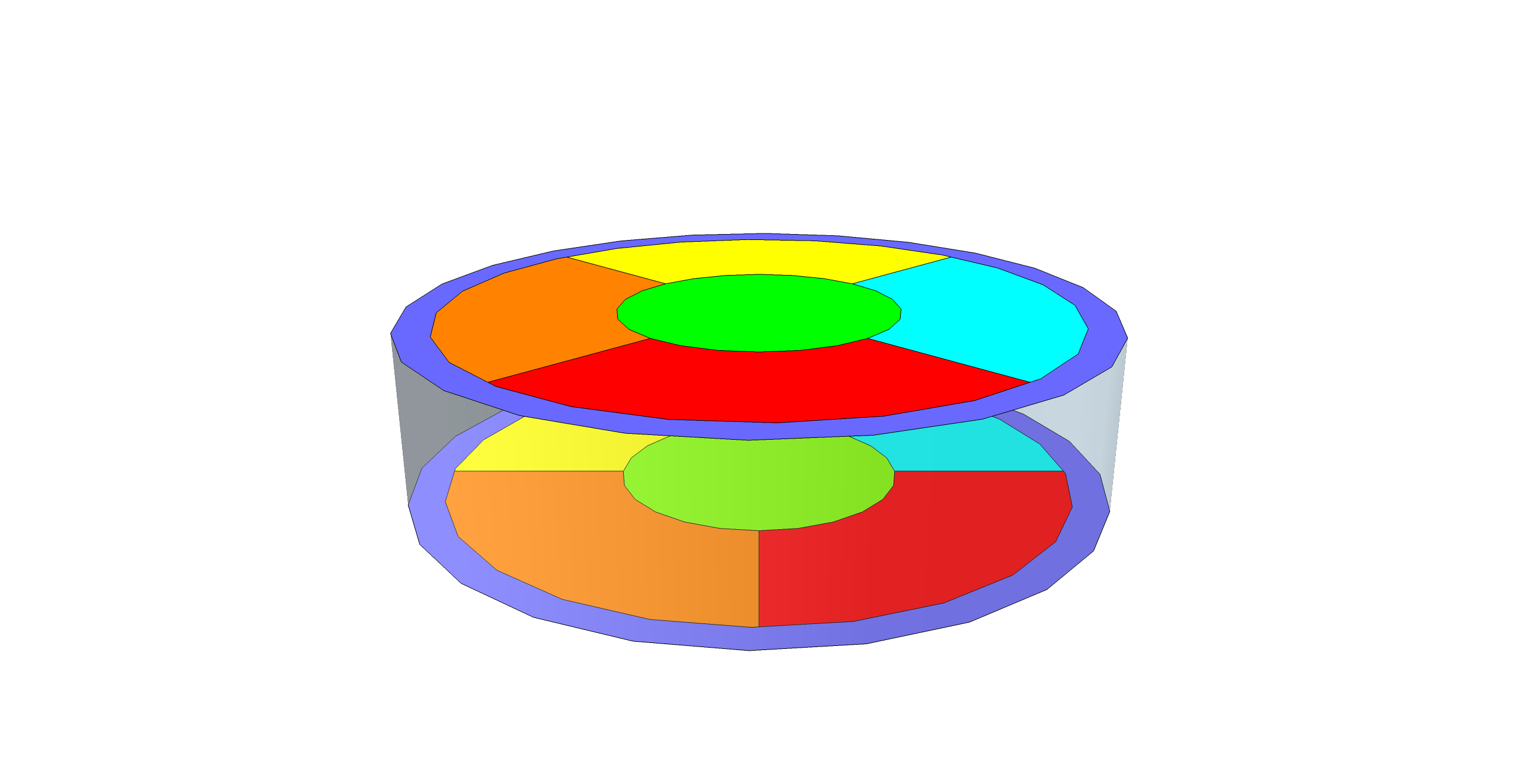}}
\caption{Channel layout for the HV (top) and iZIP (bottom) detectors. The HV detector has six phonon channels on each side, arranged as an inner ``core,'' surrounded by three wedge shaped channels and two outer rings designed to reject events near the edge. Each channel contains hundreds of lithographically defined superconducting sensors. The wedge channels on the bottom surface are rotated by 60$^\circ$ with respect to those on the top.
The interleaved Z-sensitive Ionization Phonon (iZIP) detector also has six phonon channels on each side, arranged as an inner core, surrounded by four wedge shaped channels and one outer ring. An ``outer'' ionization channel shares the same area and is interleaved with the outermost phonon ring, and an ``inner'' ionization channel is interleaved with the remaining phonon channels. The wedge channels on the bottom surface are rotated by 45$^\circ$ with respect to those on the top.}
\label{fig:DetectorSchematic}
\end{figure}

SuperCDMS SNOLAB will include a mixture of detectors composed of silicon (Si) and germanium (Ge), providing complementarity especially in the search for sub-GeV dark matter. 
These detectors consist of cylindrical crystals, 100\,mm in diameter and 33.3\,mm thick. Each Ge(Si) crystal has a mass of 1.39(0.61)\,kg. Two detector designs, denoted HV and iZIP, have common physical dimensions and are fabricated from the same materials using the same techniques. Details of the superconducting sensors, patterned lithographically on the top and bottom surfaces, and the operating bias voltages differentiate a detector as an HV or an iZIP detector.  
The HV detectors are designed to have better sensitivity for mass $\lesssim 5\GeVcc$~\cite{Agnese:2015nto}, while the iZIPs will have better sensitivity above $\sim 5\GeVcc$ because of their capability to discriminate between electron-recoil (ER) and nuclear-recoil (NR) interactions~\cite{Agnese:2014PRL}. 

HV detectors have six phonon sensors on each face with no ionization sensors, as shown in the top image in Fig.~\ref{fig:DetectorSchematic}. The phonon-only sensor layout allows for better phonon collection and thus a better phonon energy resolution and detector sensitivity at lower recoil energies than a similar iZIP detector~\cite{Kurinsky:2016ichep}. The HV detectors are intended to be operated at a bias of up to $\sim$100~V. This bias makes it possible to take advantage of the Luke-Neganov effect~\cite{Luke:1988JAP,Neganov:1985OI} to increase the phonon signal by the amount of work performed by the electric field on the charge carriers produced by an interaction as they move across the detector.
The phonon amplification can boost low-energy events above the operating threshold of the phonon sensors (which  depends only on the phonon sensor properties), effectively further increasing the detector's sensitivity to low-energy recoils. Without the ability to identify the type of recoil from an interaction, the data from the HV detectors will be dominated by ER backgrounds. The Luke-Neganov effect, however, decreases the relative rate of ER backgrounds with respect to NRs in a given energy range. The effect is due to the increased amount of ionization for ER with respect to NR interactions. ERs from a given energy bin are therefore spread out over a larger ``total phonon'' range, effectively decreasing the background rate per keV for low-mass dark matter particles~\cite{Pyle:2012LTD14}.

On the iZIP detectors (bottom image in Fig.~\ref{fig:DetectorSchematic}), the top and bottom surfaces are each instrumented with six phonon sensors interleaved with an inner and an outer ionization sensor. iZIP detectors provide the ability to distinguish beta- or gamma-induced ERs from neutron- or dark matter-induced NRs through the ``yield,'' the ratio of ionization production to recoil energy. This yield-based discrimination effectively removes all ER backgrounds in the bulk of the detector above a few keV. 

A voltage bias of $\sim$5--10\,V is applied to the ionization channels, while the interleaved phonon channels are grounded. The resulting field structure, along with the sensor layout, provides excellent rejection of surface backgrounds~\cite{Agnese:2013APL}. This ability to reject the vast majority of backgrounds on an event by event basis means the iZIP detectors can be operated in a nearly background-free mode.

The HV and iZIP detector technologies are complementary to one another and provide, respectively, access to lower-energy recoils without recoil-type discrimination and at slightly higher recoil energies the ability to discriminate the primary recoil type. The overlap in sensitivity for dark matter masses in the range of 1--10\GeVcc for the two detector technologies is expected to provide an excellent opportunity for assessing the backgrounds of the experiment and cross-checking any potential candidate dark matter signals.

The initial payload, detector performance and anticipated total exposures for the SNOLAB experiment are summarized in Table~\ref{tab:ExpOperations}. The detectors will be deployed in four towers of six detectors each. 
\begin{table}[htp]
\begin{tabular}{  l  c  c  c  c } \hline
&\multicolumn{2}{c}{iZIP} & \multicolumn{2}{c}{HV}\\ 
&  Ge & Si    & Ge & Si \\ \hline
Number of detectors & 10 & 2 & 8 & 4\\
Total exposure (kg\(\cdot\)yr) & 56  & 4.8 & 44 & 9.6 \\
Phonon resolution (eV) & 50 & 25 & 10 & 5\\
Ionization resolution (eV) & 100 & 110 & -- & -- \\
Voltage Bias (V) & 6 & 8 & 100 & 100\\ \hline
\end{tabular}

\caption{The anticipated, exposures and detector parameters for the SuperCDMS SNOLAB experiment. The exposures are based on 5 years of operation (from 2020--2024) with an 80\% live time. The quoted phonon energy resolutions represent the r.m.s.\ values of the total measured quantity (\textit{i.e.}, combining all active sensors). The quoted ionization resolution is derived from the readout electronics equivalent noise charge value of 33\textit{e} and represents the r.m.s.\ energy resolution of a single channel for electron recoils.}
\label{tab:ExpOperations}
\end{table}

\subsection{Cryostat and Shielding}
\label{sec:SNOBOX}

\begin{figure*}[htbp!]
\begin{center}
\includegraphics[width=\textwidth]{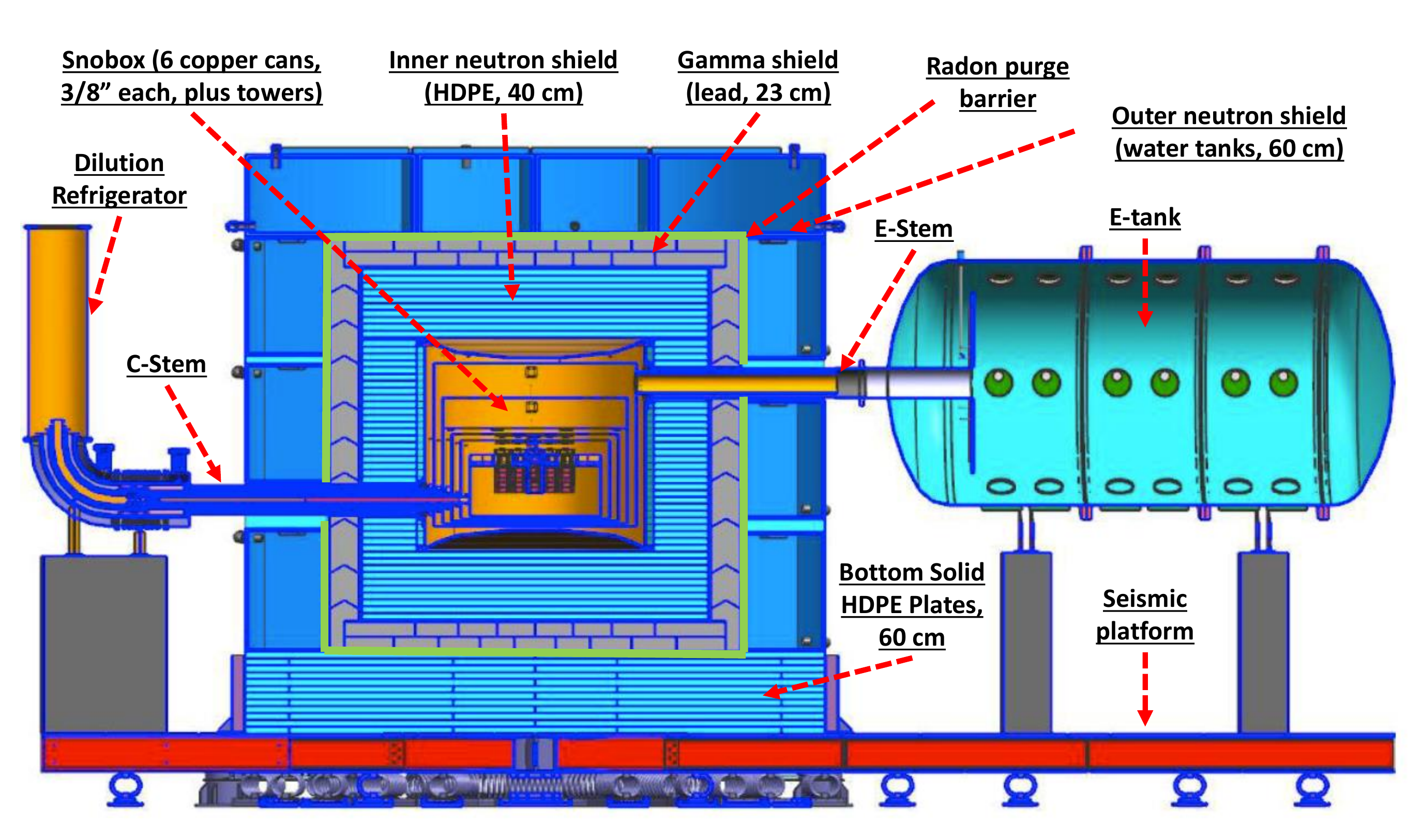}
\end{center}
\caption{A schematic of the experiment shield and cryostat layers. The assembly rests on top of a seismic platform to provide isolation from seismic events. The outer water tanks provide protection from cavern neutrons. A gamma shield protects from external gamma-rays and the inner polyethylene layers serve to absorb radiogenic neutrons emitted from the cryostat and gamma shield.  }
\label{fig:snolab}
\end{figure*}

The detector towers will be cooled to $\sim$15--30\,mK using a dilution refrigerator that utilizes cryocoolers to establish 50 and 4\,K thermal stages. The cold region of the full experiment is referred to as the SNOBOX. 
As shown in Fig.~\ref{fig:snolab}, the SNOBOX consists of six cylindrical copper cans suspended by Kevlar ropes. Each SNOBOX can is mapped onto a thermal stage of the refrigerator. The outermost can along with the stems and E-tank form the vacuum system. 

A 40\,cm thick layer of polyethylene surrounds the SNOBOX and serves to moderate and absorb neutrons produced by radiogenic contamination, as well as provides shielding from external neutrons. This inner polyethylene layer is surrounded by a 23\,cm thick gamma shield made from low-activity lead. 
The lead shield layer is surrounded by a thin metal shield to block Rn diffusion into the inner shielding layers. This volume will be purged with boil-off nitrogen gas to reduce the overall Rn levels and the backgrounds caused by prompt Rn daughters.   
The outermost shield layer consists of polyethylene and water tanks that provide additional shielding from the cavern neutron flux. 

The shield and SNOBOX are penetrated in two locations opposite each other, as shown in Fig.~\ref{fig:snolab}. The electronics stem (E-stem) provides a path for twisted-pair cables to run between the cold hardware and the electronics tank, which forms the vacuum bulkhead where signals emerge. The cryogenics stem (C-stem) connects the various layers of the SNOBOX to external cryogenic systems. 

%%
%% BACKGROUND SOURCES
%%

\section{Background Sources}
\label{sec:Backgrounds}

In this section, we describe two broad categories of background sources anticipated for the SuperCDMS SNOLAB experiment: (1) sources that produce energy depositions throughout the detector crystal volume and (2) sources that produce energy depositions primarily on or very near the surfaces of the detector crystal.  These categories are further divided into ER (\textit{e.g.},\ betas or Compton scatters) and NR events. Each of these event types must be tracked separately for each type of detector because of different detector response functions, fiducial efficiencies, and analysis efficiencies, as discussed in Section~\ref{sec:Detector}.

\subsection{Bulk Event Background Sources}
\label{sec:volumesources}
The background sources described in this subsection can produce events that occur throughout the detector volume (``bulk'' events).

\subsubsection{Detector Contamination}
The dominant backgrounds expected for the HV detectors are due to radioactive impurities within the detector crystals. Table~\ref{tab:detectorcontamination} presents the detector contamination levels assumed in the calculation of the SuperCDMS SNOLAB sensitivities shown in Fig.~\ref{fig:OI-Limits}.

\begin{table}[htp]
\begin{tabular}{ c c c c c c} \hline
 &         & Production Rate & \multicolumn{2}{c}{Concentration}\\
 &  & (atoms/kg/day)  & \multicolumn{2}{c}{(decays/kg/day)} \\
Material & Isotope &   & HV & iZIP \\ \hline
Ge & $^{3}$H & 80  & 0.7 & 1.5 \\
Si & $^{3}$H & 125 & 1 & 2 \\

Si & $^{32}$Si & --  & 80 & 80\\ \hline
\end{tabular}

\caption{Assumptions used to determine the $^{3}$H and $^{32}$Si detector contamination levels for the SuperCDMS SNOLAB sensitivities shown in Fig.~\protect{\ref{fig:OI-Limits}}.  The assumed sea-level cosmic-ray exposure for the HV(iZIP) detectors is 60(125) days, followed by a 365 day underground ``cooldown'' period before acquisition of science data. $^{32}$Si is intrinsic to the production process and is expected to be the same for iZIP and HV detectors. }
\label{tab:detectorcontamination}
\end{table}

\paragraph{Cosmogenically produced $^{3}$H:}
For both the Ge and Si detector crystals, exposure to high-energy cosmic-ray secondaries (\textit{i.e.}, neutrons, protons, and muons) results in the production of tritium ($^{3}$H) as a spallation product from interaction of the cosmic-ray secondaries with the nuclei in the detector crystals~\cite{Avignone:1992jg}. The long half-life of tritium ($t_{1/2} = 12.3$\,years) results in an accumulation of this radioactive impurity whose \mbox{$\beta$-decay} product has an endpoint energy of 18.6\,keV. The tritium background is modeled using a generic $\beta$-decay energy spectrum~\cite{Neary:1971}, the production rates and activation times shown in Table~\ref{tab:detectorcontamination} for tritium in Ge~\cite{Armengaud:2016aoz} and in Si~\cite{tritium:paper}, and the detector crystal masses given in Table~\ref{tab:ExpOperations}.

\paragraph{Naturally occurring $^{32}$Si:}
This radioactive isotope is produced as a spallation product from cosmic-ray secondaries on argon in the atmosphere~\cite{Lal:1960}. The $^{32}$Si atoms make their way into the terrestrial environment through aqueous transport (\textit{i.e.}, rain and surface water). Consequently, the exact source and location of the silicon used in the production and fabrication of silicon detectors may impact the concentration level of $^{32}$Si observed in future detectors. The long $\sim$153\,year half-life of $^{32}$Si~\cite{Ouellet:2011nds} means the concentration of $^{32}$Si measured in decays/kg/day is essentially fixed once the Si detector crystal is grown.  Although the literature~\cite{plaga:1991} suggests low-radioactivity silicon is in principle available for rare-event searches, we assumed the central value of the $^{32}$Si concentration recently measured by the DAMIC collaboration in their CCD detectors, which was \(80^{+110}_{-65}\)\,decays/kg/day at a 95\% confidence level~\cite{Aguilar-Arevalo:2015lvd}. The $^{32}$Si background is modeled using a generic $\beta$-decay energy spectrum~\cite{Neary:1971}.

\paragraph{Ge activation lines:}
Exposure of the Ge detector substrates to high-energy cosmic-ray secondaries results in the production of several radioisotopes that decay by electron capture. We include here the eight isotopes observed in the CoGeNT experiment~\cite{Aalseth:2010vx} that are sufficiently long-lived to contribute background in the SuperCDMS SNOLAB Ge detectors: $^{68}$Ge and $^{68}$Ga daughter, $^{65}$Zn, $^{73}$As, $^{57}$Co, $^{55}$Fe, $^{54}$Mn and $^{49}$V. Each decay can proceed via electron capture from the K, L or M shell, giving rise to a total of 24 spectral peaks (\textit{cf.}~Fig.~\ref{fig:HVBackgroundSpectraRaw}). 
We scale the K-shell peak rates by the ratio of the sea-level exposure for the SuperCDMS detectors and the reference CoGeNT detector, and the rates of the L- and M-shell lines are scaled according to their relative branching fractions.

\subsubsection{Material Activation}
Exposure to high-energy cosmic-ray secondaries results in the production of long-lived radioisotopes in the construction materials surrounding the detectors. In particular, the cosmogenic activation of copper presents a background source for the SuperCDMS SNOLAB experiment. Copper is used both for the detector tower mechanical assembly and the nested cylindrical cryostat canisters. Table~\ref{tab:copperexposure} presents the assumptions used to assess the emission rates due to cosmic-ray activation of these components.
\begin{table}[htp]
\begin{tabular}{c c c c} \hline
        & \multicolumn{1}{c}{Production Rate} & \multicolumn{2}{c}{Contamination Rate ($\mu$Bq/kg)}\\
Isotope & (atoms/kg/day)  &  Housings/Towers & Cryostat\\ \hline
$^{46}$Sc & 4.6  & 0.88 & 0.62\\
$^{48}$V  & 9.5  & 0.76 & 0.25\\
$^{54}$Mn & 19   & 7.9 & 12\\
$^{56}$Co & 20   & 3.5 & 2.3\\
$^{57}$Co & 155  & 62 & 89\\
$^{58}$Co & 143  & 23 & 13\\
$^{59}$Fe & 39   & 2.9 &  0.9\\
$^{60}$Co & 181  & 47 & 90 \\ \hline
\end{tabular}
\caption{Assumptions used to determine the cosmogenic exposure and activation of copper for the SuperCDMS SNOLAB sensitivities shown in Fig.~\protect{\ref{fig:OI-Limits}}.  The sea-level activation rates are taken from \cite{Cebrian:2010ApP}, except for $^{48}$V, which is taken from~\cite{Laubenstein2009}. A sea-level exposure of 90~days is assumed for copper in the detector housings and towers, followed by a 90~day underground ``cooldown'' period before acquisition of science data. The sea-level exposure and underground-cooldown periods for the copper cryostat cans are both assumed to be 180\,days. At the time of this publication only $^{57}$Co, $^{58}$Co, $^{60}$Co, and $^{54}$Mn have been simulated for the tower and housing copper; the decay rate for the other listed isotopes is at least 5$\times$ lower. For the cryostat cans, only $^{60}$Co is presently included; emissions from the other isotopes are lower in energy and thus less penetrating and can be neglected.}
\label{tab:copperexposure}
\end{table}

\subsubsection{Material Contamination}
\label{sec:materialcontamination}
Radioactive impurities are introduced in all materials at some level during the manufacturing process.  The $^{238}$U and $^{232}$Th isotopes are unstable but long-lived and are present in most materials at low concentrations. Both of these isotopes have a chain of decay daughters that are assumed to be in secular equilibrium. Additionally, isotopes such as $^{40}$K and $^{60}$Co are naturally present in many materials because of their long half-lives, but they do not have accompanying series of daughter radioisotopes. Table~\ref{tab:radioimpurities} lists the assumed levels of radioactive impurities present in materials used in the construction of the SuperCDMS SNOLAB experiment. Cirlex is used to fashion the clamps that hold the detectors in place within the copper housings and the detector electronics readout PCBs. Additionally, for each of the ``FETCard,'' ``SquidCard,'' ``TowerTruss'' and flex cable volumes shown in Fig.~\ref{fig:towergeo}, we assign a total emission of 0.1\,mBq for $^{238}$U, $^{232}$Th, and $^{40}$K to account for all of the individual small components (resistors, etc.).  

\begin{table}[htp]
\begin{tabular}{l c c c c c c} \hline
              & \multicolumn{5}{c}{Impurity concentrations (mBq/kg)} & Ref. \\
Material      & $^{238}$U & $^{232}$Th & $^{40}$K & $^{60}$Co & $^{137}$Cs & \\
\hline
Copper        &    0.07   &     0.02    &   0.04   &   *   &      &  \cite{Aprile:11ap}\\
Cirlex        &    6.3    &     2.2     &    1.6   &   0.01    & 0.01    & \cite{Loach:2016fsk} \\
Kevlar        &    430    &     140     &    870   &         &        & \cite{Arpesella2002}\\
$\mu$-metal   &    4.2    &     4.2     &    1.7   &   0.51    & 0.27   & \cite{Loach:2016fsk} \\
HDPE          &    0.6    &     1.5     &    1.9   &   0.13    & 0.19     & \cite{Loach:2016fsk} \\
Lead          &    0.66   &     0.5     &     7    &         &      & \cite{Aprile:11ap}\\
Polypropylene &    0.6    &     1.5     &    1.9   &   0.13    & 0.19     & HDPE \\
Water         &    0.6    &     1.5     &    1.9   &   0.13    & 0.19     & HDPE \\ \hline
\end{tabular}
\caption{Radioactive impurity concentrations assumed for construction materials contained within the SuperCDMS SNOLAB experiment. $^{60}$Co in copper (indicated by *) is assumed to be produced cosmogenically rather than introduced during production and so the assumed rate is different for different pieces; see Table~\protect{\ref{tab:copperexposure}}.  Values found for HDPE were also assumed for both polypropylene and water, which we expect to be conservative. 
}
\label{tab:radioimpurities}
\end{table}

\subsubsection{Non-Line-of-Sight Surfaces}
\label{sec:nonlossurfaces}
Materials accumulate concentrations of radioactive isotopes on surfaces exposed to air containing dust and radon. Airborne dust typically contains relatively high concentrations of $^{238}$U, $^{232}$Th, and $^{40}$K. Daughters from the decay in air of $^{222}$Rn may implant shallowly into a material surface, resulting in a buildup of the long-lived $^{210}$Pb that later decays through a short chain and produces a roughly constant emission rate of x-rays, betas and alphas.

We consider separately surfaces with and without a clear line of sight to the detector; surfaces with line of sight are generally of much greater concern and are discussed in Section~\ref{sec:surfacesources}. For surfaces without line of sight, we are concerned primarily with gamma- and x-ray emission, and to a lesser extent neutron emission, as those are the only radiation types capable of reaching the detectors. $^{210}$Pb produces some soft x-rays that may reach the detectors if emitted from surfaces very near to the detectors, and its daughter $^{210}$Bi has a moderately high energy $\beta$-decay that may in turn produce bremsstrahlung x-rays that may be more penetrating.  Finally, the alpha produced by the subsequent $^{210}$Po decay may produce neutrons via an $(\alpha,n)$ reaction on $^{13}$C, and so $^{210}$Pb accumulation on hydrocarbon surfaces such as polyethylene is a potential concern. For this analysis we assumed a $^{210}$Pb activity of 850~nBq/cm$^2$ for non-line-of-sight surfaces inside the cryostat, roughly corresponding to 100 days exposure to air with radon concentrations of 10 Bq/m$^3$.  For the outer cryostat and shielding surfaces we assume an activity of 11,000~nBq/cm$^2$ corresponding to 100 day exposure to 130~Bq/m$^3$ air. Dust has not been included at this time, but preliminary estimates indicate that it should contribute less than or comparably to background from non-line-of-sight $^{210}$Pb surface contamination.

\subsubsection{Cavern Environment}
The cavern environment background sources include naturally occurring radioactivity in the underground environment leading to gamma-rays or neutrons that potentially pass through the SuperCDMS shield and interact in the detectors.
The experiment cavern is surrounded by Norite rock that has been coated with a layer
of shotcrete. The cavern floor is concrete. The wall and floor layers have variable thicknesses
but are on the order of a few inches thick~\cite{Duncan:2010zz}.

\paragraph{Gamma-rays}
The gamma-ray background is modeled as a $^{40}$K decay along with decay chains in secular equilibrium for $^{238}$U and $^{232}$Th. 
Those gamma-ray emission spectra were simulated and evaluated for their leakage through the shielding, in particular the E- and C-stem penetrations (\textit{cf.}\ Fig.~\ref{fig:snolab}).
The gamma-ray flux was estimated using results from assays of rocks collected in the SNOLAB
ladder labs and Monte Carlo simulations. This source, however, is not included in the present model because of limited simulation statistics and because preliminary results indicated that it is subdominant to other sources.
The gamma shield is being further optimized so that this source should have a negligible impact on the total background budget. 

\paragraph{Neutrons}
The neutron background from the cavern environment is modeled as two components: neutrons from $^{238}$U spontaneous fission and neutrons produced through $(\alpha,n)$ reactions in the rock due to U- and Th-chain alpha-emitting isotopes. 
The shape of the neutron spectra are calculated using a modified version of {\sc Sources 4c}, which calculates neutron spectra for spontaneous fission and the material specific $(\alpha,n)$ process in Norite and shotcrete~\cite{sources4,sources4c:IDM}. The neutrons are propagated through the materials using Monte Carlo simulations.
 The overall normalization of the spectrum is taken from Ref.~\cite{Smith:2012fq}, which specifies a flux of 4000 fast neutrons/(day$\cdot \mathrm{m}^2$).

\paragraph{Radon}
Radon decays in the mine air produce moderately high-energy gamma-rays via the $^{214}$Pb and $^{214}$Bi daughters. Decays occurring outside the shield contribute to the total gamma-ray flux already considered for the cavern as a whole and are not considered separately.  If air in the region between the lead gamma shield and the SNOBOX were allowed to mix freely with the mine air, radon decays in this region would produce a significant background.  However, this contribution is assumed to be made negligible by the hermetic radon purge surrounding the gamma shield. 

\subsubsection{{\it In situ} Cosmic Ray Induced Backgrounds}
The overburden provided by SNOLAB significantly reduces, but does not eliminate, cosmic rays. 
Muons may pass directly through a detector or create secondary particles through interactions with the surrounding materials; high-energy neutrons produced via spallation are our primary concern.  We simulated muons with the angular and energy distribution appropriate for SNOLAB depth parameterized by Mei and Hime~\cite{Mei:2006PRD} from a $\sim$10~meter diameter plane. 
Unlike SuperCDMS Soudan, the detector at SNOLAB will not have a muon veto. 

\subsubsection{Coherent Neutrino Interactions}
Although not expected to be a significant background for the initial SuperCDMS SNOLAB experimental program, the interaction of solar neutrinos through coherent elastic scattering off detector nuclei currently presents a limiting background source to future low-mass dark matter search experiments~\cite{Billard:2013qya}. The decay of $^{8}$B at the end of the pp-III solar fusion reaction chain produces the primary solar neutrino background for future expansions of the SuperCDMS SNOLAB experiment. The background is estimated using a theoretical value for the solar neutrino fluxes~\cite{Bahcall:2004pz}, the theoretical $^{8}$B solar neutrino energy spectrum~\cite{Strigari:2009bq}, and the coherent elastic neutrino-nucleus scattering cross-section~\cite{Freedman:1977xn}.

\subsection{Surface Event Background Sources}
\label{sec:surfacesources}
The second broad category of background sources produces energy depositions on or very near the surfaces of the detector substrates. These backgrounds are explicitly related to the exposure of the detectors and their housing materials (primarily Cu) to $^{222}$Rn and its progeny during fabrication, testing and installation. When radon decays in air, for example, its daughters can plate-out onto a surface and the subsequent Po alpha-decays can cause the long-lived $^{210}$Pb daughter to become implanted into the surface. If the implantation occurs far enough along in the fabrication process, it is no longer practical to remove and is thus a source of background that will be present for the duration of the experiment. Other surface contaminants are possible (\textit{e.g.},~$^{14}$C or $^{39}$Ar) but have not been identified in SuperCDMS detectors and are not considered here.

The $^{210}$Pb decay chain produces a variety of radiation types that are generally not very penetrating. Consequently, aside from the few exceptions pointed out in Section~\ref{sec:nonlossurfaces} above, $^{210}$Pb surface contamination is a background concern only if there is a clear line of sight (\textit{i.e.}, no intervening material) between the location of the contamination and a detector surface. There are three principle radioisotopes in the decay chain: $^{210}$Pb, $^{210}$Bi and $^{210}$Po.  $^{210}$Pb yields a combination of low-energy betas and \mbox{x-rays}, resulting in a near-surface ER background in all detector types, whose spectrum resembles a $\beta$-decay spectrum with an $\sim$60\,keV endpoint but superimposed with several x-ray lines (most notably at 46.5\,keV). $^{210}$Bi undergoes $\beta$-decay with an $\sim$1.2\,MeV endpoint, also resulting in a near-surface ER background but with a harder spectrum and no lines. Finally, $^{210}$Po decays by emitting an alpha so energetic that it is generally outside the dark matter signal region. Unlike in the preceding two decays, however, the $^{206}$Pb daughter nucleus recoils with sufficient energy to potentially create a NR in the dark matter signal region.  If the decay occurs on a detector surface such that the $^{206}$Pb recoil is directed into the surface, the full 103\,keV recoil energy is deposited in the detector.  If the decay occurs on a nearby surface (\textit{e.g.}, detector housing), the energy of the $^{206}$Pb nucleus may be degraded because of the implantation depth of the $^{210}$Po parent, yielding a continuum of NR energies up to $\sim$100\,keV.

The detector response and detector type are particularly important considerations when evaluating the impact of these surface backgrounds on the experimental sensitivity. $^{206}$Pb recoils incident on a detector face will predominantly be tagged by a large energy deposition in the adjacent detector from the associated alpha, and thus such events will not contribute to the background of a dark matter search. Similarly, $^{210}$Pb decays often result in simultaneous energy depositions in adjacent detectors, allowing them to be rejected as dark matter candidates.  As discussed in~\cite{Agnese:2013APL}, interleaving the iZIP phonon and ionization sensors enables discrimination of surface events at the detector faces. Sensor modularity enables fiducialization of the signal to reject surface events incident at the sidewalls~\cite{Agnese:2014PRL}. This ``radial'' fiducialization is expected to be effective for HV detectors as well as iZIPs. The SuperCDMS SNOLAB detector response is discussed in more detail below in Section~\ref{sec:Detector}.

For the evaluation of the SuperCDMS SNOLAB sensitivities shown in Fig.~\ref{fig:OI-Limits}, we assume a total $^{210}$Pb surface activity of 50\,nBq/cm$^2$ for line-of-sight surfaces (the detector surfaces and inner surfaces of the copper housings). This is the same level of activity that was observed on the surfaces of the Ge iZIP detectors in the SuperCDMS Soudan experiment, inferred from the rate \textit{versus} time of 5.3\,MeV $^{210}$Po alphas incident on the detector faces.

%%
%% SIMULATION
%%

\section{Monte Carlo Simulation}
\label{sec:Simulation}

{\it SuperSim} is a package used to simulate background sources for SuperCDMS.  
The {\it SuperSim} package is a set of detector models and applications that have been written in C++ on top of the {\sc Geant4} framework utilizing the {\sc Shielding} physics list~\cite{Geant_applications,Geant_toolkit}. The {\sc Shielding} physics list is a reference physics list provided by the Geant4 collaboration and is recommended for underground physics and shielding applications~\cite{shielding}. 

\begin{figure}[htp]
\includegraphics[width=0.9\columnwidth]{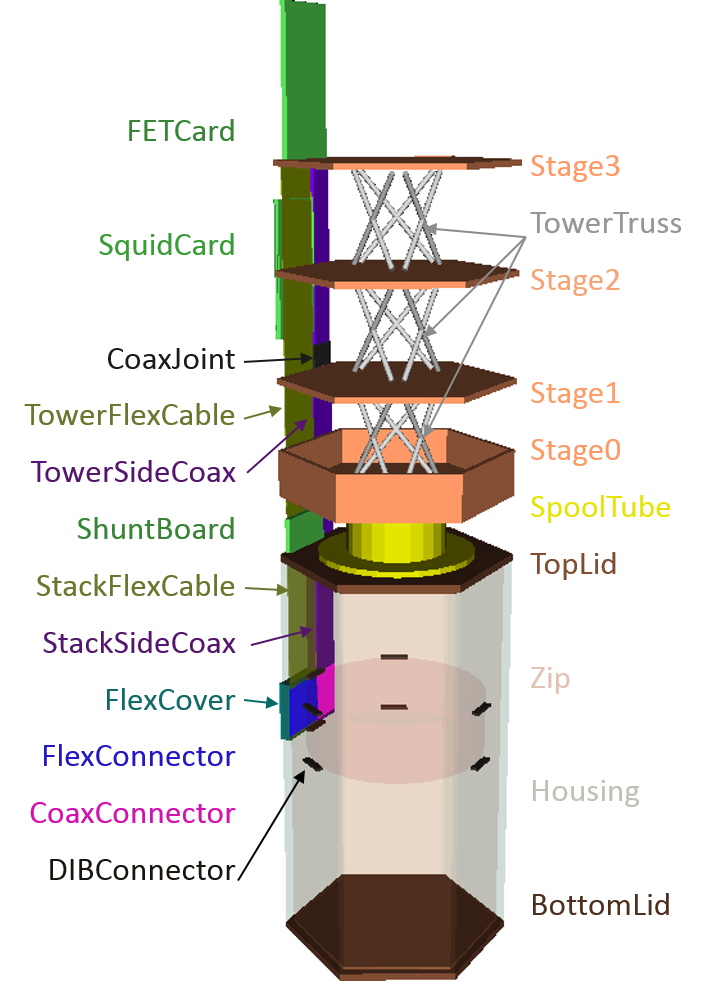}
\caption{Detailed simulation geometry of the SuperCDMS SNOLAB tower mechanical support and instrumentation for one of six detectors in a tower.}
\label{fig:towergeo}
\end{figure}

One of the primary functions of {\it SuperSim} is to provide {\sc Geant4} with the geometry of the shielding and inner array of towers as portrayed in Fig.~\ref{fig:snolab}. The nested cryostat geometry is reproduced in {\it SuperSim} along with the approximate geometries for the tower components. The {\it SuperSim} model includes the detector housings, tower assembly, and the locations of electronic components.  Most of the background rate predictions discussed here are based on simulations of an older proposed design consisting of seven total towers in a larger cryostat. A few dedicated studies indicate that normalized event rates are not significantly affected by this difference. 
Figure~\ref{fig:towergeo} shows the tower mechanical support and the instrumentation for a single detector (HV or iZIP). This level of fidelity within the {\it SuperSim} geometry model is used for evaluation of the background from small, discrete electrical components.

Primary event generation can occur through several methods. Isotopes can be specified and the decay will be handled by the physics processes in {\sc Geant4}. In some simple cases, such as gamma-ray emission, the dominant characteristic gamma-rays are coded directly  into {\it SuperSim}. Primaries can also be thrown from a user-defined spectrum. For U- and Th-chain gamma decays,  gamma-ray spectra assuming secular equilibrium have been calculated and are used to generate those primaries. Spectra for neutrons are calculated using the modified version of the {\sc Sources 4c} software package. 

The majority of radiogenic background simulations are produced by contaminating volumes with primaries. Contamination can be bulk, surface or restricted to a portion of a volume. Generally, the primary locations are uniformly distributed within the specified region, but an exponential surface depth can be specified for surface sources. The angular distribution of decays can be either isotropic or biased in order to improve simulation efficiency. 

The location, total energy deposited, and interaction channel (NR or ER) are recorded for every interaction in the detectors.  No detector response effects, including phonon and ionization generation and propagation, are simulated. Energy depositions in separate detectors within 1~ms are bunched into ``triggers'' for the purpose of identifying multiple scatters. A single {\sc Geant4} event may produce multiple ``triggers,'' especially when simulating primaries leading to radioactive decay chains. 

The probability of double scattering of a dark matter particle is effectively zero. Many of the background sources, however, have a significant probability to interact with multiple detectors. Thus, requiring that only one detector have an interaction per simulated event serves to discriminate against background sources. In practice, it is necessary to set an energy threshold in order to avoid an unreasonable rejection of events due to detector noise. This threshold, which is applied to the total observed phonon signal, corresponds to seven times the expected detector resolution $\sigma_{Ph}$ (\textit{cf.}~Table~\ref{tab:ExpOperations}), and is 350(175)\,eV for Ge(Si) iZIP detectors. The threshold for the HV Ge(Si) detectors was similarly set at 70(35)\,eV. If a ``trigger'' has an energy deposition in one detector and the energy deposited in the other detectors is below the threshold, the trigger is classified as a single-scatter event. If two or more detectors have energy deposited that exceeds the energy threshold, the trigger is classified as a multiple-scatter and will be rejected by the analysis-level event selection.

Further steps are required to convert numbers of events from {\it SuperSim} into a rate ($R$) of expected background events for each energy range of interest as illustrated by Equation~\ref{eq:R}.
\begin{equation}
\begin{split}
R \left[\mathrm{\frac{counts}{kg\cdot keV\cdot yr}}\right] = & C \left[\mathrm{\frac{counts}{primary}}\right] \times \frac{1}{M\Delta E [\mathrm{kg\cdot keV}]} \\
&\times R_d \left[\mathrm{\frac{decays}{year}}\right] \times F_p \left[\mathrm{\frac{primaries}{decay}}\right]
\end{split}
\label{eq:R}
\end{equation}

Here $C$ is the number of events passing a set of selection criteria for a given number of input primaries.
The second term of this equation normalizes the rate to units of kg$\cdot$keV for each target mass and energy range of interest.  The third term, $R_d$, is the number of decays per year for the given background source. The final term is $F_p$, which encodes the number of primaries per decay and is typically unity.

\begin{figure*}[htp!]
\includegraphics[width=0.475\textwidth, trim={0 0 50 0}, clip]{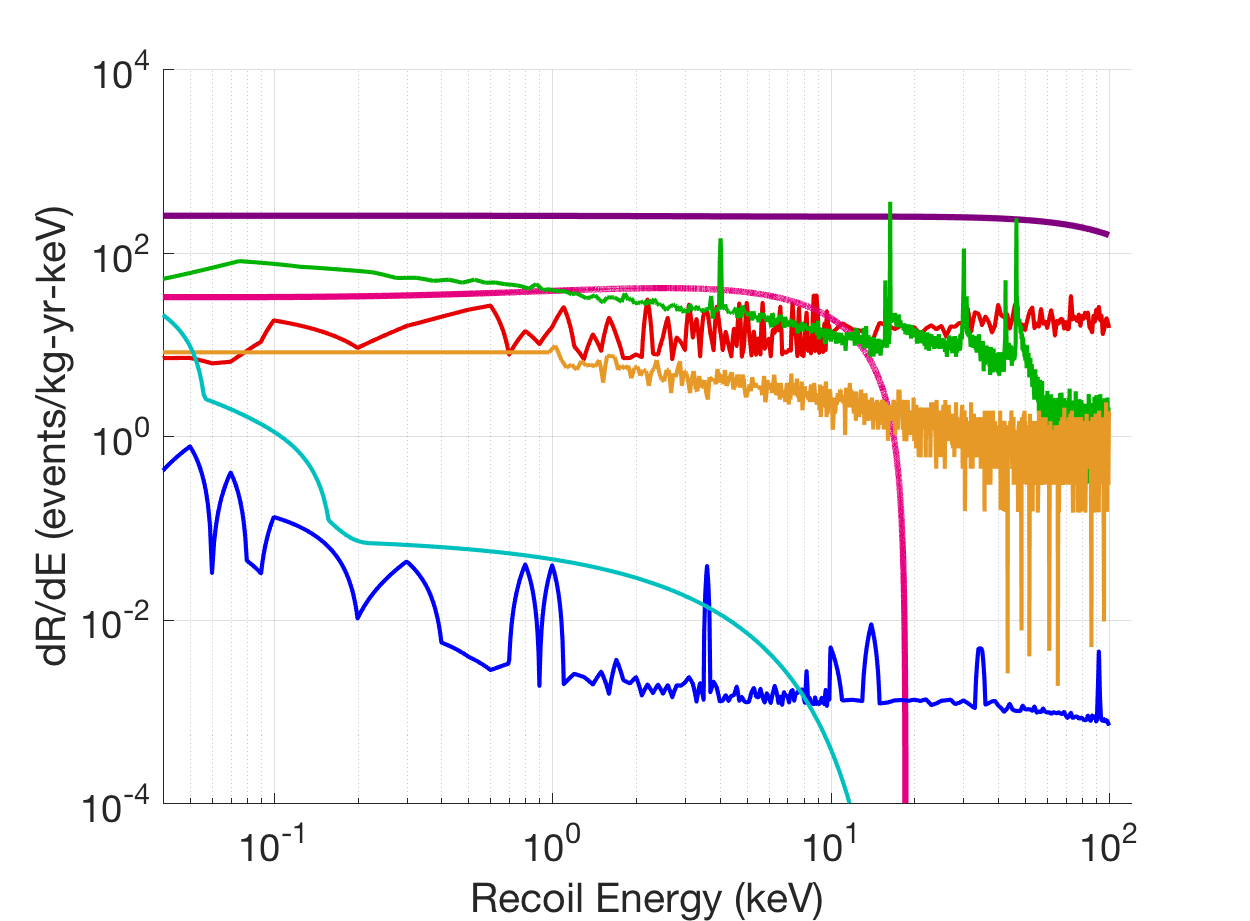}
\includegraphics[width=0.475\textwidth, trim={0 0 50 0}, clip]{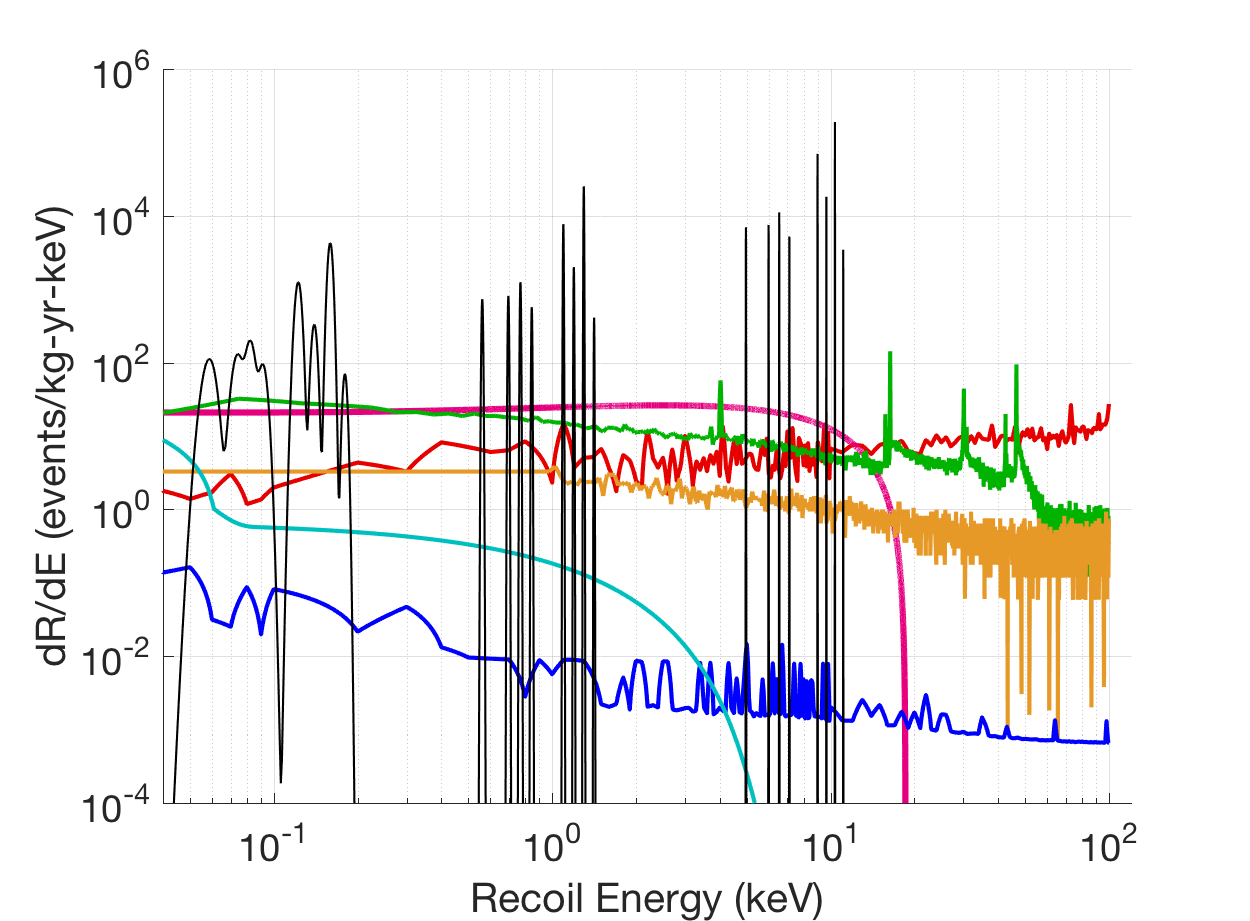}
\caption{Raw background spectra of single scatter interactions in a Si (left) and Ge (right) detector obtained from the Monte Carlo simulation. The spectra are broken out by components and shown as a function of recoil energy (ER or NR depending on the interaction). $^3$H (pink) and $^{32}$Si (purple) are the largest individual contributors to the backgrounds in the Ge and Si detectors, respectively. The Ge activation lines (black) are shown convolved with a 10\,eV r.m.s.\ resolution (\(\sigma_{Ph}\) for the Ge HV detectors) to allow them to be clearly displayed in this figure. The remaining components are Compton scatters from gamma-rays (red), surface betas (green), surface \(^{206}\)Pb recoils (orange), neutrons (blue) and coherent elastic neutrino-nucleus scattering (cyan). Note that the neutron spectrum (blue) has some spurious structure from the limited simulation statistics in the cavern component of the neutron background (\textit{cf.} Table~\ref{tab:backgroundbudget}).}
\label{fig:HVBackgroundSpectraRaw}
\end{figure*}

%%%
%%% BACKGROUND BUDGET TABLE HERE
%%%

\begin{table*}[htp!]
\begin{tabularx}{0.85\textwidth}{l | X X X X | X X} 
\hline
\multicolumn{1}{c|}{{\bf ``Singles'' Background Rates}} & \multicolumn{4}{c|}{Electron Recoil} & \multicolumn{2}{c}{Nuclear Recoil ($\times10^{-6}$)}  \\
\multicolumn{1}{c|}{{\bf (counts/kg/keV/year)}} & \multicolumn{1}{c}{Ge HV} & \multicolumn{1}{c}{Si HV} & \multicolumn{1}{c}{Ge iZIP} & \multicolumn{1}{c|}{Si iZIP} & \multicolumn{1}{c}{Ge iZIP}  &\multicolumn{1}{c}{Si iZIP}  \\
\hline
Coherent Neutrinos & & & & & 2300. & 1600. \\
Detector-Bulk Contamination & $\phantom{00}$21. & $\phantom{0}$290. & $\phantom{000}$8.5 & $\phantom{0}$260. &  &  \\
Material Activation & $\phantom{000}$1.0 & $\phantom{000}$2.5 & $\phantom{000}$1.9 & $\phantom{00}$15. & & \\
Non-Line-of-Sight Surfaces & $\phantom{000}$0.00 & $\phantom{000}$0.03 & $\phantom{000}$0.01 & $\phantom{000}$0.07 & $\phantom{000}$-- & $\phantom{000}$-- \\
Bulk Material Contamination & $\phantom{000}$5.4 & $\phantom{00}$14. & $\phantom{00}$12. & $\phantom{00}$88. & $\phantom{0}$440. & $\phantom{0}$660. \\
Cavern Environment & $\phantom{000}$-- & $\phantom{000}$-- & $\phantom{000}$-- & $\phantom{000}$-- & $\phantom{0}$510. & $\phantom{0}$530. \\
Cosmogenic Neutrons & & & & & $\phantom{00}$73. & $\phantom{00}$77.\\
\hline
Total & $\phantom{00}$27. & $\phantom{0}$300. & $\phantom{00}$22. & $\phantom{0}$370. & 3300. & 2900. \\
\hline
\end{tabularx}
\caption{Summary of the background rates for detector-bulk single scatter events in the SuperCDMS SNOLAB dark matter experiment. The rates shown are prior to any detector response considerations, averaged over 3\,eV to 2\,keV for HV detectors and 1--50\,keV for iZIP detectors. Note that these energy ranges are meant to roughly represent the detectors' range, but are not used in calculating the search sensitivity. Cells marked as ``--'' will have nonzero contributions but are believed to be small and have not been included in the current model. }
\label{tab:backgroundbudget}
\end{table*}

For contaminant sources, $R_d$ is the volume mass times the contamination rate. 
The expression for $R_d$ for cosmogenic sources is shown in Equation~\ref{eq:cosmo} below.
\begin{equation}
\begin{split}
R_d(\mathrm{cosmogenic})& = m_c \cdot r_{act} \cdot \left[1-\exp\left(-\frac{t_{act}}{\tau}\right)\right] \\
& \times \left[\exp\left(-\frac{t_{cool}}{\tau}\right) - \exp\left(-\frac{t_{cool}+t_{run}}{\tau}\right)\right] \\ 
&\times \frac{\tau}{t_{run}}
\end{split}
\label{eq:cosmo}
\end{equation}

The situation is more complicated for cosmogenic sources because there is an exposure period ($t_{act}$), a cooldown period before acquisition of science data begins ($t_{cool}$), and the subsequent decay of the contaminant over the course of actual running ($t_{run}$). 

Figure~\ref{fig:HVBackgroundSpectraRaw} shows the single scatter spectra, separated into component backgrounds, expected in the Si and Ge detectors. Table~\ref{tab:backgroundbudget} provides further details on the relative contributions of different sources to the gamma-ray and neutron backgrounds over an assumed energy range.

%%
%% RESPONSE
%%

\section{Detector Response}
\label{sec:Detector}

We use an analytic framework to model the detector response and calculate sensitivity projections. This framework takes as input the background spectra in Fig.~\ref{fig:HVBackgroundSpectraRaw} obtained from the Monte Carlo simulation described in Section~\ref{sec:Simulation}, properties of the specific detectors considered, such as sensor resolution and bias voltage, and the exposure of the experiment to determine the response of the detectors to both the background and a hypothetical dark matter signal. The detector model also takes into account the position-dependent differences in response for both electron and nuclear recoils to calculate the energy-dependent efficiency for various cuts. The expected background after application of analysis cuts is then obtained. This is used in conjunction with the expected experiment exposure to determine the projected sensitivity. This procedure is performed separately for the Si and Ge varieties of the HV and iZIP detectors.

\subsection{Ionization and Phonon Signals}

To model the ionization and phonon production of an event in the detector, we consider two classes of events: ER and NR. For the same recoil energy $E_R$ deposited in the crystal, ERs generate a larger amount of ionization than NRs. The ratio of the ionization produced by NRs to that of ERs is called the ionization yield. 
The amount of ionization produced in an event is given by $y E_R / \epsilon$, where $y$ is the ionization yield and $\epsilon$ is the average energy required to create an electron-hole pair (taken to be 3.0\,eV in Ge and 3.82\,eV in Si). The model treats the ionization as a continuous variable, with no statistical fluctuations in the amount of produced ionization. For iZIPs, the measured ionization signal is normalized to
\begin{equation}
E_Q = \eta\, y\, E_R,
\label{eq:q}
\end{equation}
where $\eta$ is a position-dependent correction factor that accounts for lower measured ionization signals from events near the faces and sidewalls of the detector (\textit{cf.}~Table~\ref{tab:eta}).

The total phonon signal from an event, $E_{PT}$, is given by 
\begin{equation}
E_{PT} =E_R + E_{Luke} = E_R + \eta \frac{y E_R}{\epsilon} e \Delta V,
\label{eq:pt}
\end{equation}
where the first term is the recoil energy, and the second term is the additional phonon signal generated through the Luke-Neganov effect ($E_{Luke}$), with $y E_R/\epsilon$ being the amount of ionization produced, $e \Delta V$ the work done to move one charge through the crystal, and $\eta$ the same position-dependent efficiency factor described in Equation~\ref{eq:q}. For iZIPs, $E_{Luke}$ is on the order of $E_R$, but in the HV detectors, where $\Delta V$ is large, the Luke term can dominate the phonon signal.

\subsubsection{Ionization Yield in Ge}
\label{sec:yGe}
For Ge, we use an ionization yield as a function of recoil energy based on Lindhard theory~\cite{Lindhard:1963,*Lindhard:1963a,*Lindhard:1968} down to a selected energy cutoff, $E_{y_0}$, and assume it to be zero below that point. 
Experimental measurements in Ge are consistent with Lindhard theory down to at least a recoil energy of 254\,eV~\cite{Scholz:2016arXiv,Jones:1975PRA,Barker:2013ap}. We choose the cutoff \(E_{y_0}\) to be at 40\,eV, roughly a factor of 2--3$\times$ higher than the minimum energy required for dislocating a Ge atom from its lattice site~\cite{Nordlund:1998prb}.

\subsubsection{Ionization Yield in Si}
\label{sec:ySi}
\begin{figure}
\begin{center}
\includegraphics[width=0.9\columnwidth]{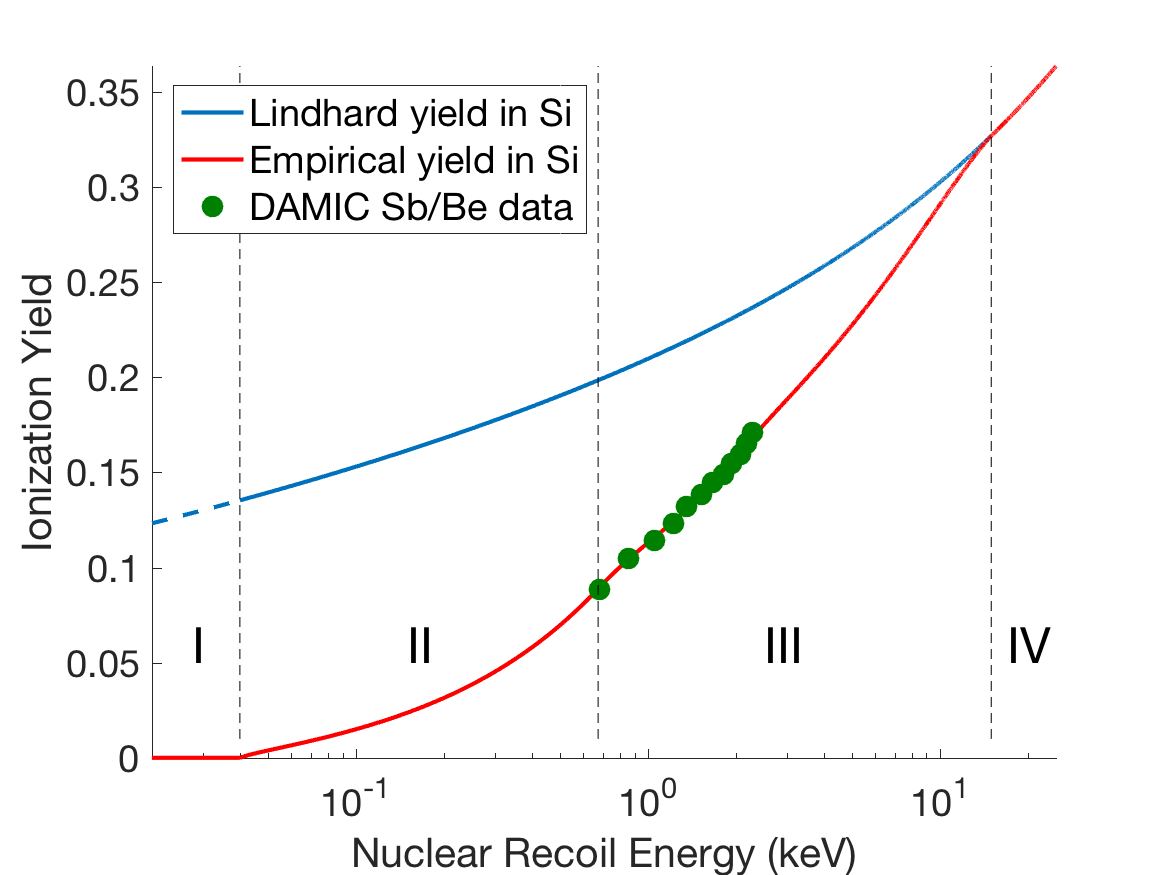}
\end{center}
\caption{Ionization yield in Si. The energy range and yield behavior of the numbered regions are:  
(I) \(0<E_{nr}<E_{y_0} (0.04\,\mathrm{keV})\), no ionization production; 
(II) \(0.04<E_{nr}<0.675\)\,keV, ionization yield described by power-law function;
(III) \(0.675<E_{nr}<15\)\,keV, ionization yield described by an empirical fit to DAMIC data~\cite{Chavarria:2016arXiv,Chavarria:2016pc}; and 
(IV) \(E_{nr}>15\)\,keV, ionization yield described by Lindhard theory.}
\label{fig:ModifiedYieldSi}
\end{figure}

For Si, we construct a yield function from a theoretical model and measured data. Above 15\,keV nuclear recoil energy, Lindhard theory is used and is consistent with experimental measurements in Si~\cite{Sattler:1965pr,Dougherty:1992pra,Gerbier:1990prd}. Between 0.675 and 15\,keV, an empirical fit to data recently published by the DAMIC experiment~\cite{Chavarria:2016arXiv,Chavarria:2016pc} is used. Below 0.675\,keV, a power law that matches the amplitude and slope of the DAMIC measurement at 0.675\,keV and decreases to zero at a selected energy cutoff $E_{y_0}$, also set to 40\,eV~\cite{Nordlund:1998prb}, is used. The yield function in Si is shown for the full energy range in Fig.~\ref{fig:ModifiedYieldSi}.

\subsubsection{Position Dependent Ionization}
In addition to the yield, which is solely a function of energy and recoil type, a correction factor \(\eta\) is applied to take into account incomplete signal collection for events near (and on) the detector faces and cylindrical sidewalls.  These correction factors are based on data obtained from the SuperCDMS Soudan experiment and test facility measurements with HV and iZIP detectors. The values of \(\eta\) and the corresponding detector-volume fractions for events in the bulk and near the surfaces for ER and NR events in Ge and Si are presented in Tables~\ref{tab:eta} and~\ref{tab:vol}; Table~\ref{tab:eta} also lists the values of $\eta$ for events \textit{on} the detector surfaces.

\begin{table}
\begin{tabular}{lcc}
\hline
Event Location and Type & Ge  & Si \\
\hline
Bulk Events & 1.0 & 1.0\\
Events near the top/bottom faces & 1.0 & 1.0\\
Events near the cylindrical sidewalls & 0.75 & 0.90\\ \hline\hline
ERs on the top/bottom faces & 0.70 & 0.65\\
ERs on the cylindrical sidewalls & 0.525 & 0.585\\
$^{206}$Pb recoils on the top/bottom faces & 0.65 & 0.65\\
$^{206}$Pb recoils on the cylindrical sidewalls & 0.488 & 0.585\\
\hline
\end{tabular}
\caption{Fraction of ionization collected by events of given locations and types.
}
\label{tab:eta}
\end{table}

\begin{table}[htp]
\begin{tabular}{lccc} \hline
					& \multicolumn{3}{c}{Volume Fraction} \\
					& \multicolumn{2}{c}{ER} & NR \\
Volume Type & Ge  & Si &  Ge \& Si\\
\hline
Bulk Events & 0.50 & 0.675 & 0.85\\
Events near the top/bottom faces & 0.056 & 0.075 & 0.05\\
Events near the cylindrical sidewalls & 0.444 & 0.25 & 0.10 \\  \hline
\end{tabular}
\caption{Fraction of a detector's volume for which ER/NR events are identified as interacting in a given volume type.}
\label{tab:vol}
\end{table}

\subsection{Background Response and Cuts}
\label{sec:Cuts}

The background components detailed in Section~\ref{sec:Backgrounds} and  Fig.~\ref{fig:HVBackgroundSpectraRaw} are classified as ER or NR. Using Equations~\ref{eq:q} and \ref{eq:pt}, the spectral response for each background component is obtained, assigning the specific $\eta$ correction factor and ionization yield as needed for the event type, location, energy, and detector material. The resulting phonon signals for the HV and iZIP detectors are shown in the left-hand spectra in Figs.~\ref{fig:HVBackgroundSpectra} and~\ref{fig:iZIPBackgroundSpectra}, respectively. Note that the energy is given in nuclear recoil equivalent energy (keVnr). In the HV detectors, ER and NR events are not differentiated.  Consequently, the HV-detector analysis is performed assuming all events are nuclear recoils, using Equation~\ref{eq:pt} to determine the equivalent $E_R$ given a measured $E_{PT}$.

For each detector type, the signal region is specified by a series of data cuts, which are applied to the left-hand spectra in Figs.~\ref{fig:HVBackgroundSpectra} and~\ref{fig:iZIPBackgroundSpectra}. The resulting signal-region background expectations are shown in the right-hand spectra in Figs.~\ref{fig:HVBackgroundSpectra} and~\ref{fig:iZIPBackgroundSpectra}. These include a 5\% reduction in overall exposure intended to model global data-quality cuts, which are energy- and position-independent, as well as energy-dependent fiducial-volume cuts. For the latter, the response of nuclear recoils in the detector bulk to a discriminant metric is modeled as having an energy-dependent mean and standard deviation, allowing the definition of an energy-dependent $2\sigma$ signal acceptance band (\textit{i.e.},\ for nuclear recoils in the bulk). The response of each background type (\textit{e.g.},\ ERs in the bulk and near/on surfaces) to the discriminant is modeled, and an energy-dependent passing efficiency is determined for each based on the separation of the metric between bulk nuclear recoils and the specific background. 

\subsubsection{HV Detector Cuts}
For the HV detectors, only a single cut is considered in addition to the data-quality cuts. It is a phonon-based radial fiducial-volume cut intended to remove background events (with reduced Luke amplification) from the outer cylindrical walls of the detectors (as demonstrated in Ref.~\cite{Agnese:2015nto}).

\subsubsection{iZIP Detector Cuts}
For iZIP detectors, which make use of the ionization signal, the fiducial volume is defined by phonon- \textit{and} ionization-based depth (or $z$-position) \textit{and} radial cuts. Additionally, an ionization yield cut that uses both the ionization and phonon information, and has excellent discrimination power for rejecting ER backgrounds in the bulk, is applied. As illustrated in Fig~\ref{fig:iZIPBackgroundSpectra}, the bulk ER backgrounds above 2\,keV are reduced by a factor of $\sim 10^6$; it is expected there will be zero ER events, in that energy range, over the full exposure of the experiment.
Below 2\,keV, the yield-based discrimination begins to degrade. The energy range from 2\,keV down to the analysis threshold, however, remains useful for the lowest-mass dark matter searches despite the nonzero background expectation. 

\subsubsection{Applying the Analysis Cuts}
\begin{figure*}[p]
\includegraphics[width=0.475\textwidth, trim={0 0 50 0}, clip]{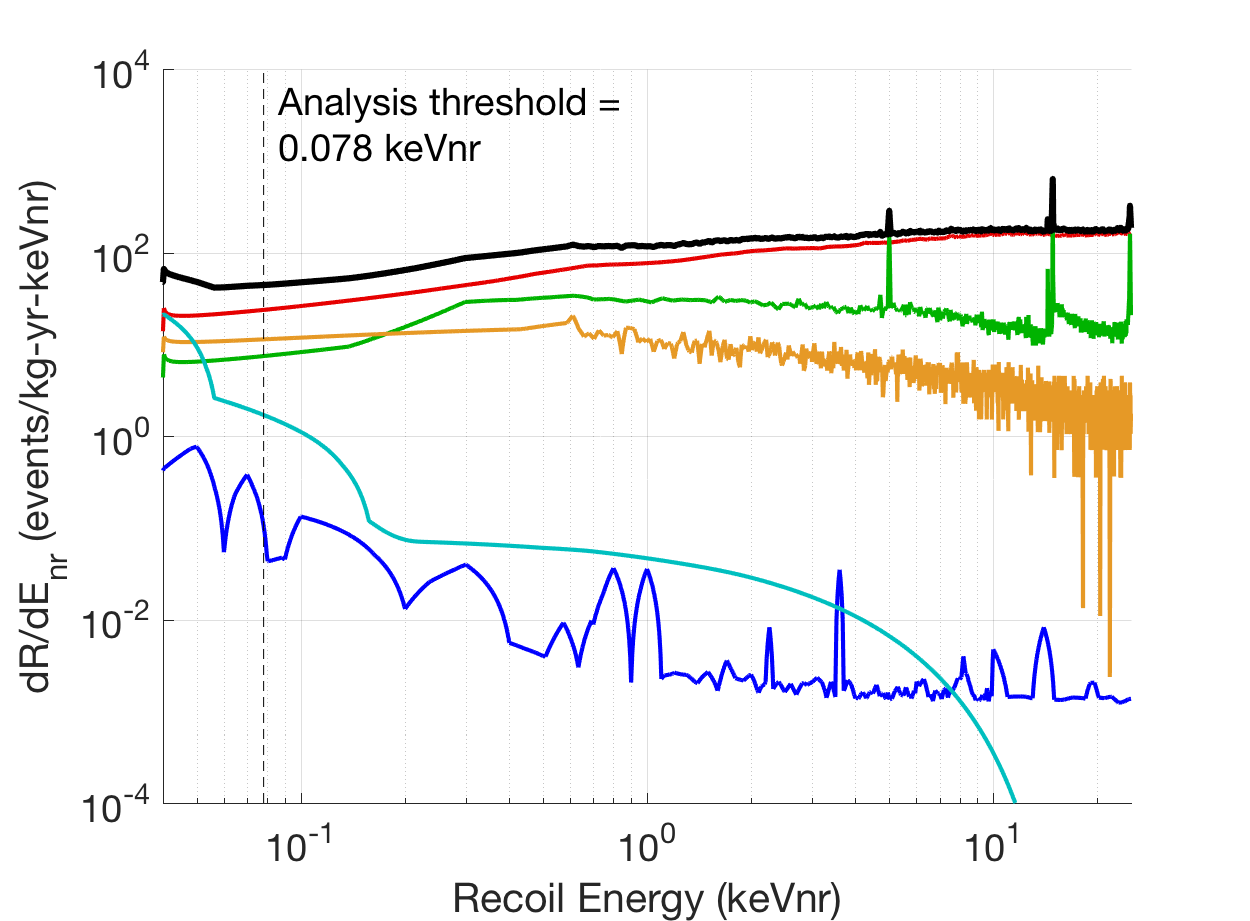}
\includegraphics[width=0.475\textwidth, trim={0 0 50 0}, clip]{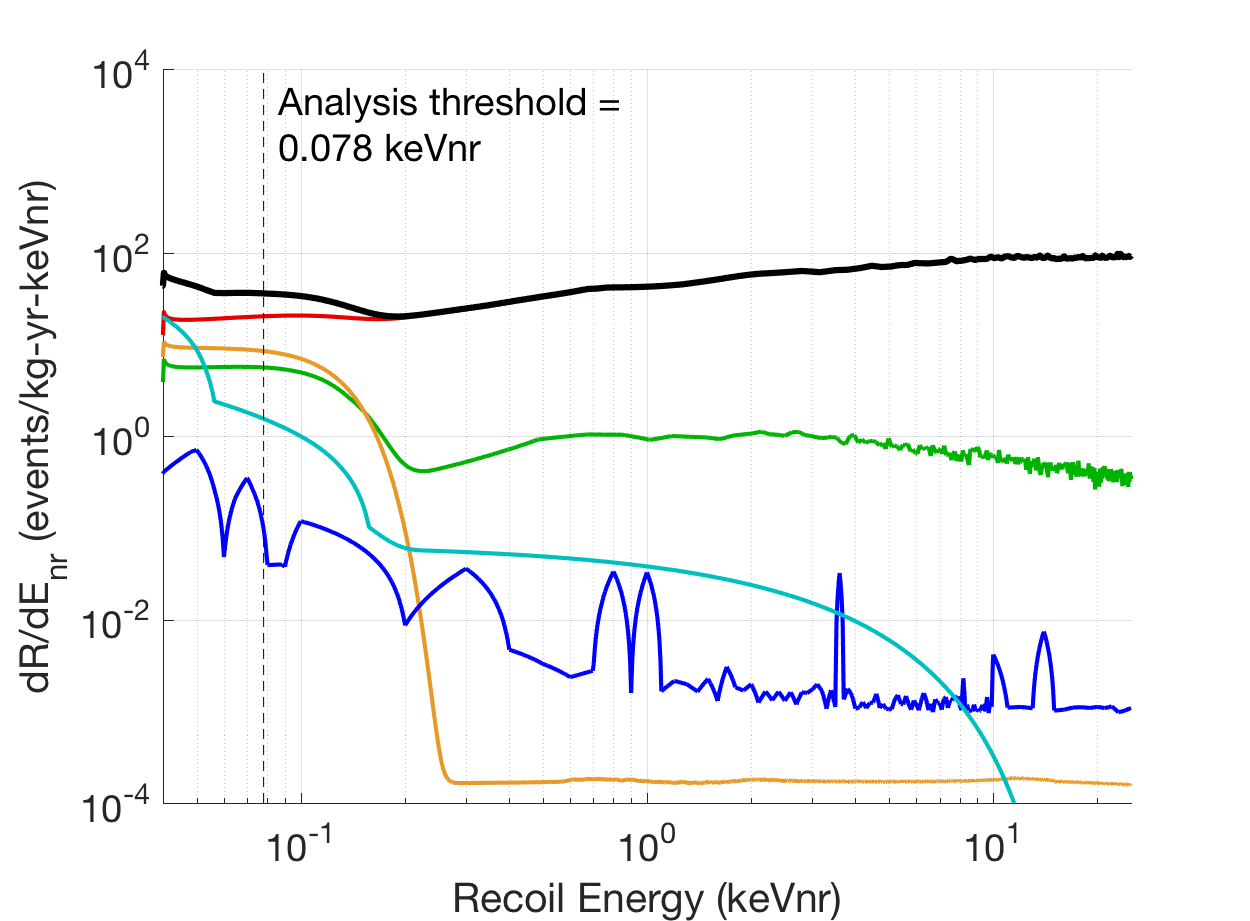}
\includegraphics[width=0.475\textwidth, trim={0 0 50 0}, clip]{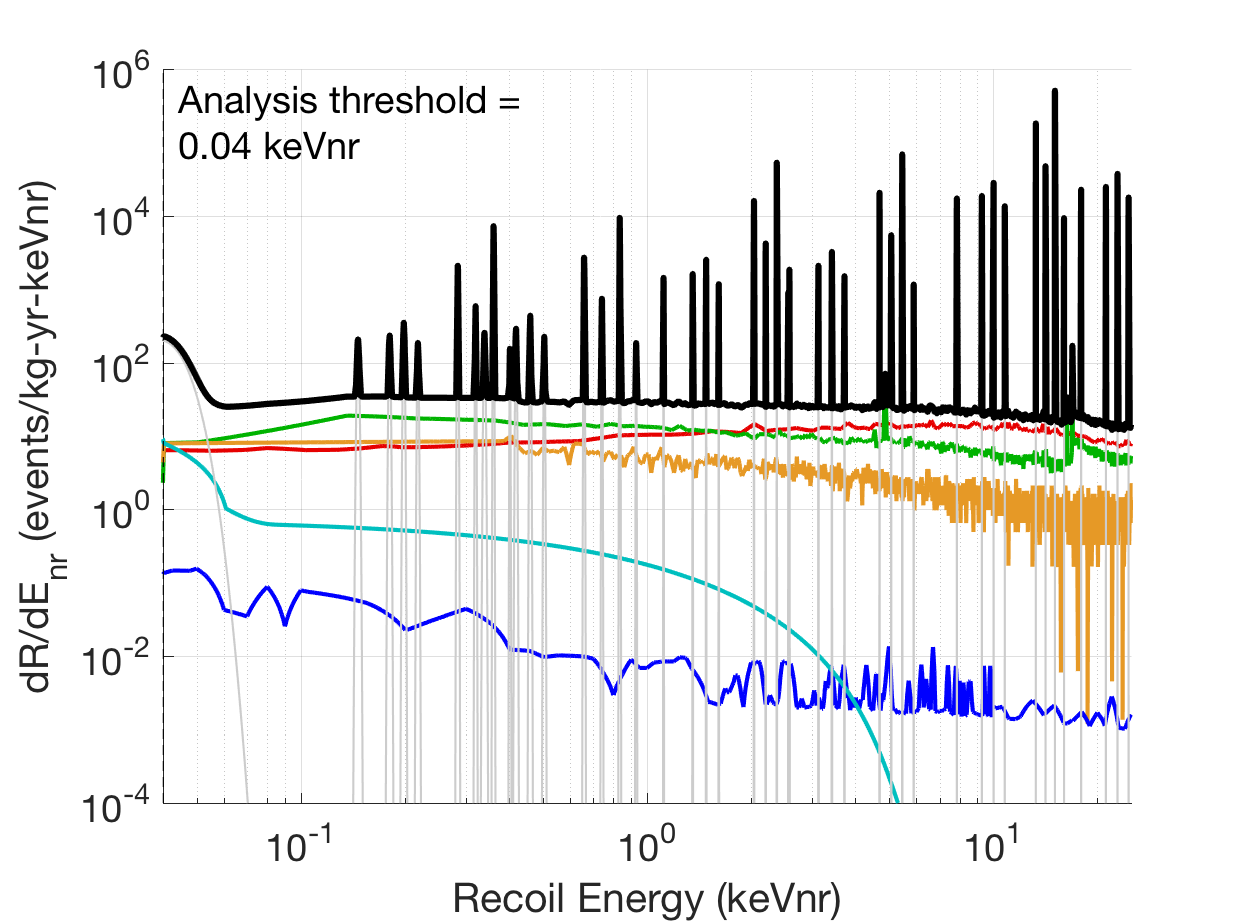}
\includegraphics[width=0.475\textwidth, trim={0 0 50 0}, clip]{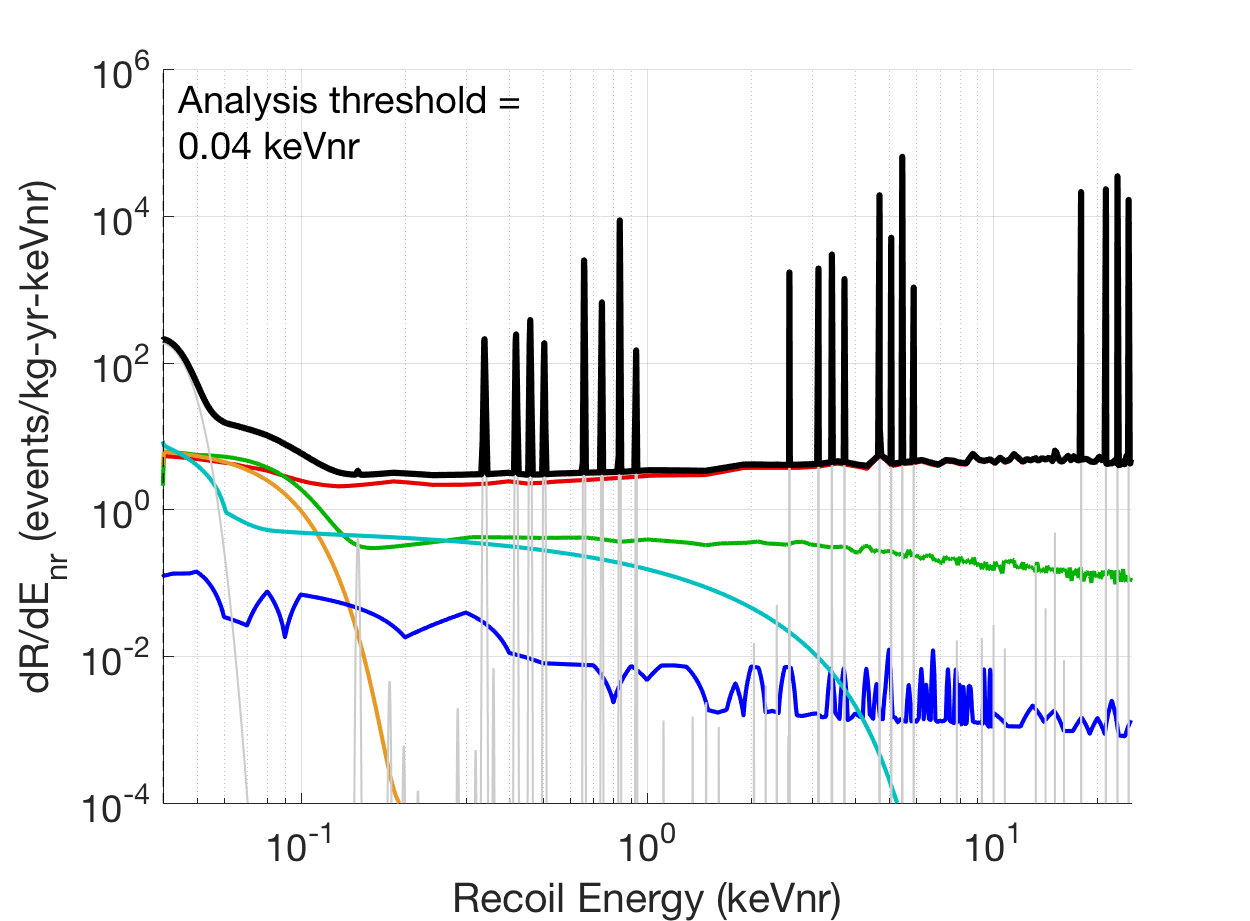}
\caption{Background spectra, before (left) and after (right) analysis cuts in Si (top) and Ge (bottom) HV detectors, shown as a function of nuclear recoil energy (keVnr). Thick black lines represent the total background rates. Electron recoils from Compton gamma-rays, $^3$H, and $^{32}$Si are grouped together (red). The Ge activation lines (grey) are shown convolved with a 10\,eV r.m.s.\ resolution. The remaining components are surface betas (green), surface \(^{206}\)Pb recoils (orange), neutrons (blue), and CE$\nu$NS (cyan). The large number of apparent Ge activation lines in the before cuts (left) spectrum is due to the discrete variation in the detector's ionization response as a function of position ($\eta$) in the simplified detector response model used in this analysis; namely a given recoil energy can be reconstructed to multiple nuclear recoil energy values, according to the discrete parameterization in Table~\ref{tab:eta}. After application of the radial fiducial-volume cut, the number of apparent lines is reduced because only the detector region with $\eta=1$ remains (minimizing the shortcomings of the simplified model). The value of $\eta$ for an actual detector is expected to be continuous, resulting in a less pronounced and more smeared out reconstruction of the activation peaks in the pre-cut spectrum compared to the model used here.}
\label{fig:HVBackgroundSpectra}
\end{figure*}

\begin{figure*}[p]
\includegraphics[width=0.45\textwidth, trim={0 0 50 0}, clip]{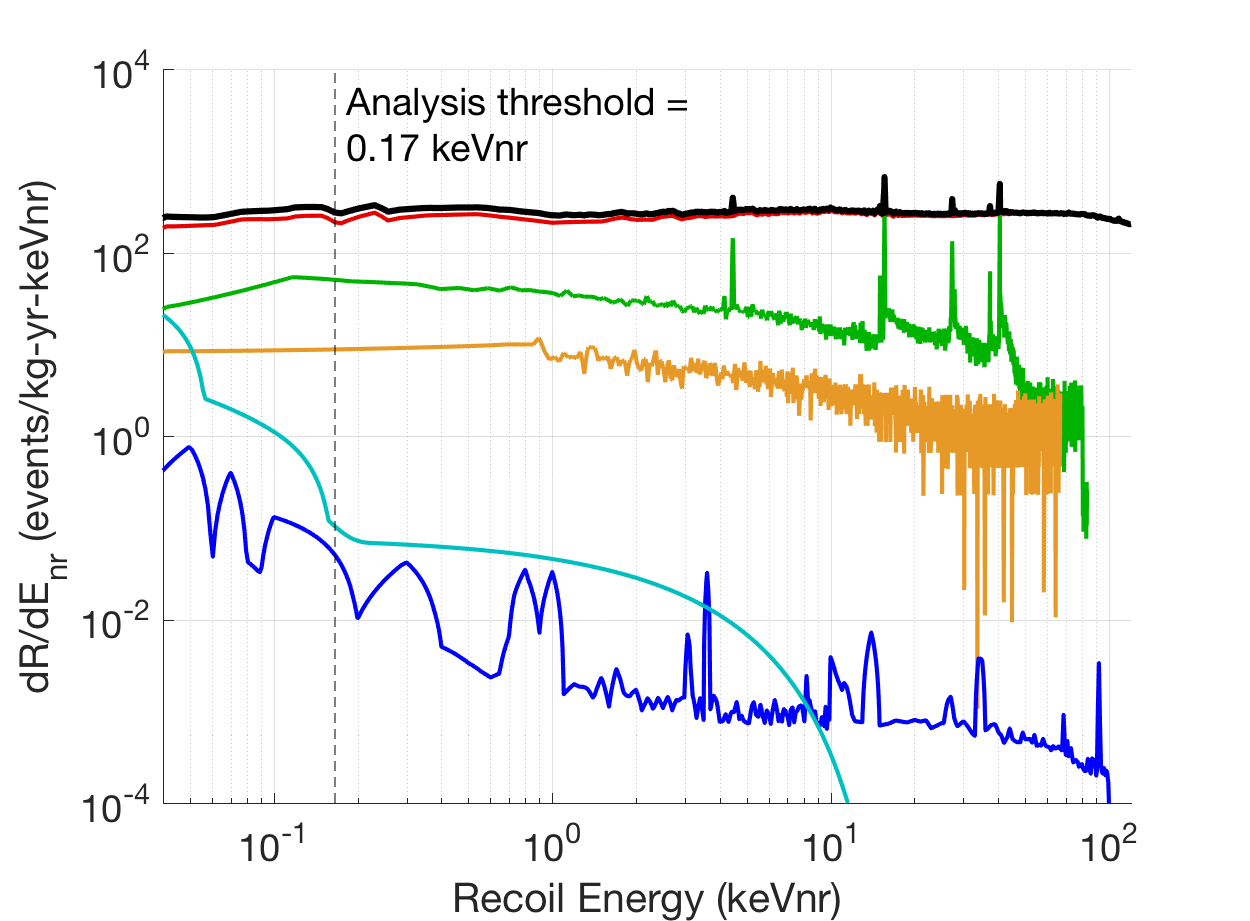}
\includegraphics[width=0.45\textwidth, trim={0 0 50 0}, clip]{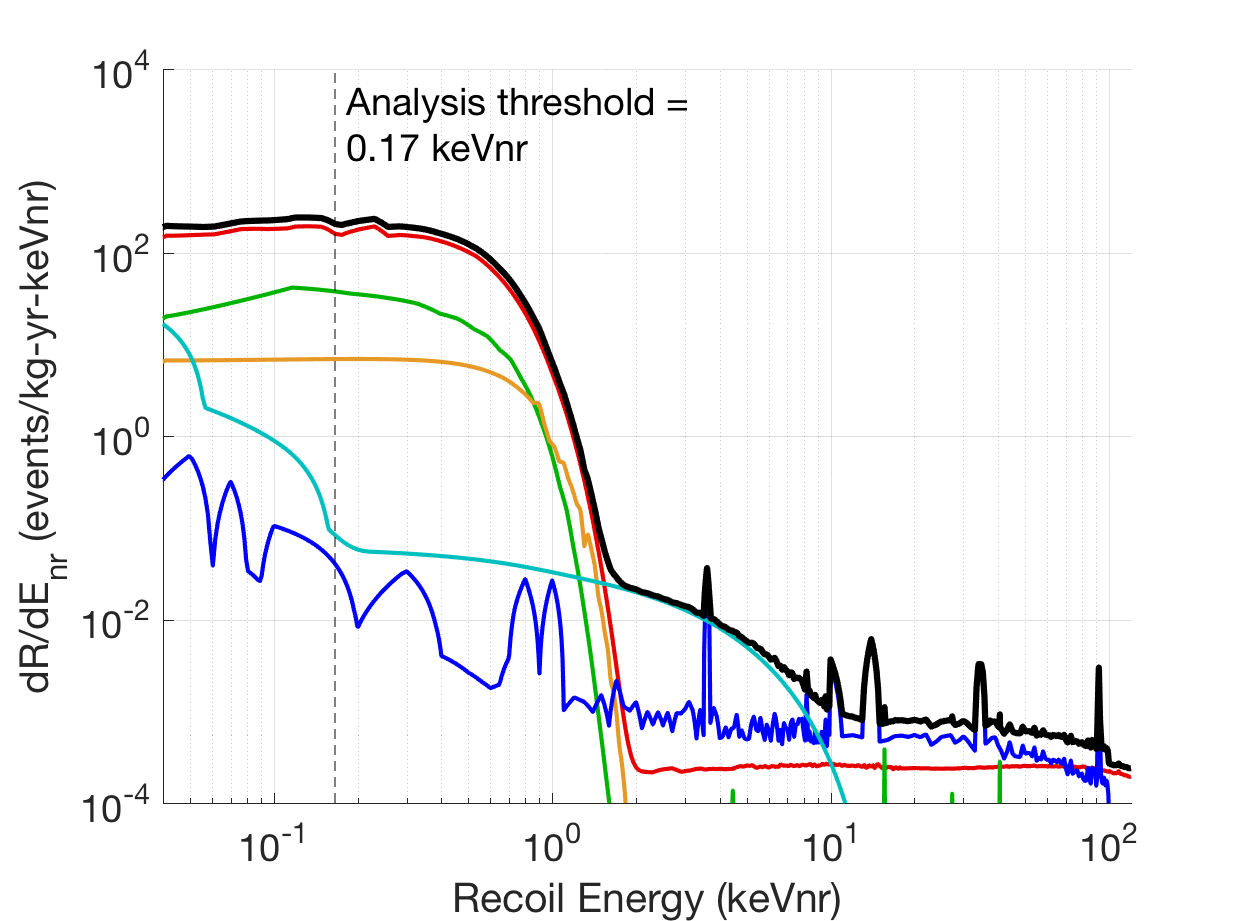}
\includegraphics[width=0.45\textwidth, trim={0 0 50 0}, clip]{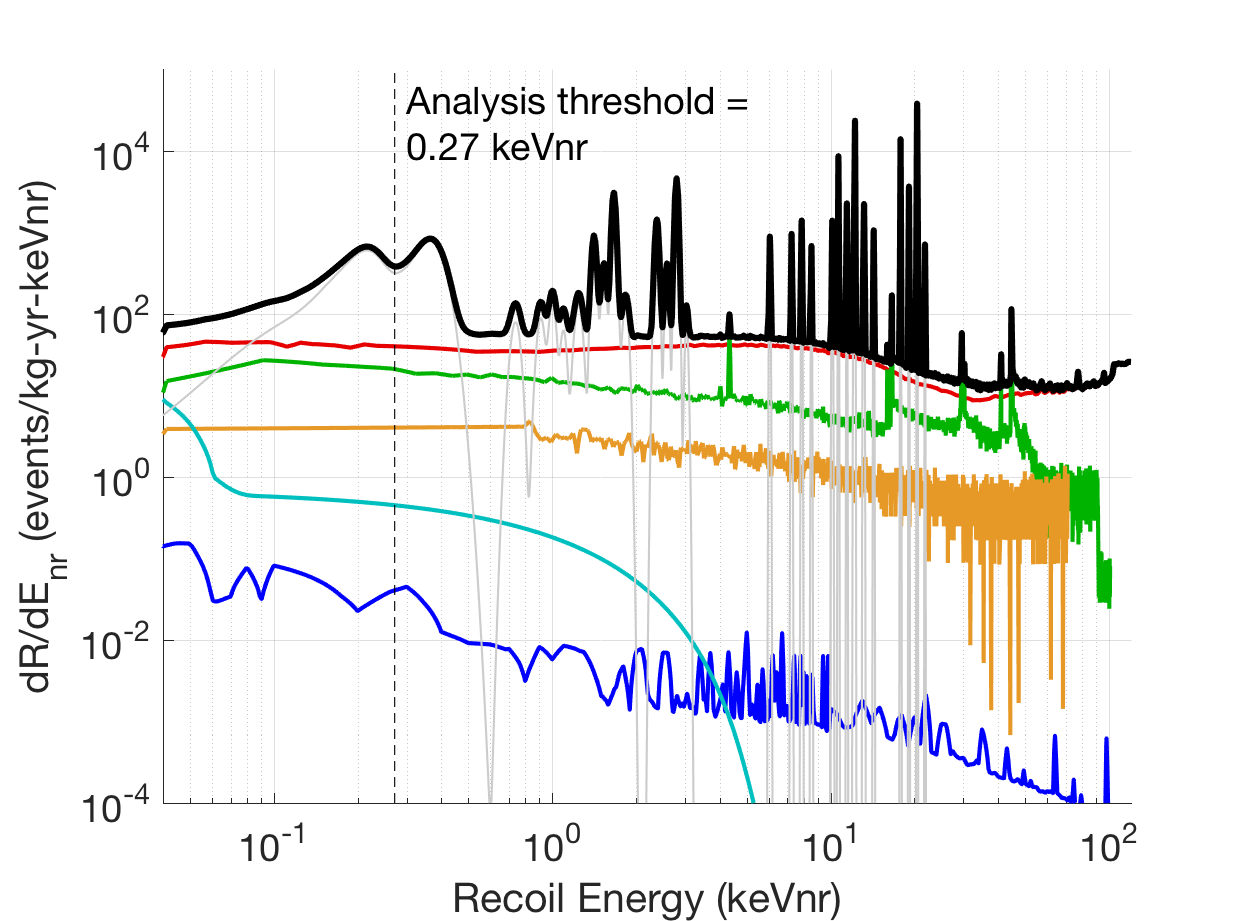}
\includegraphics[width=0.45\textwidth, trim={0 0 50 0}, clip]{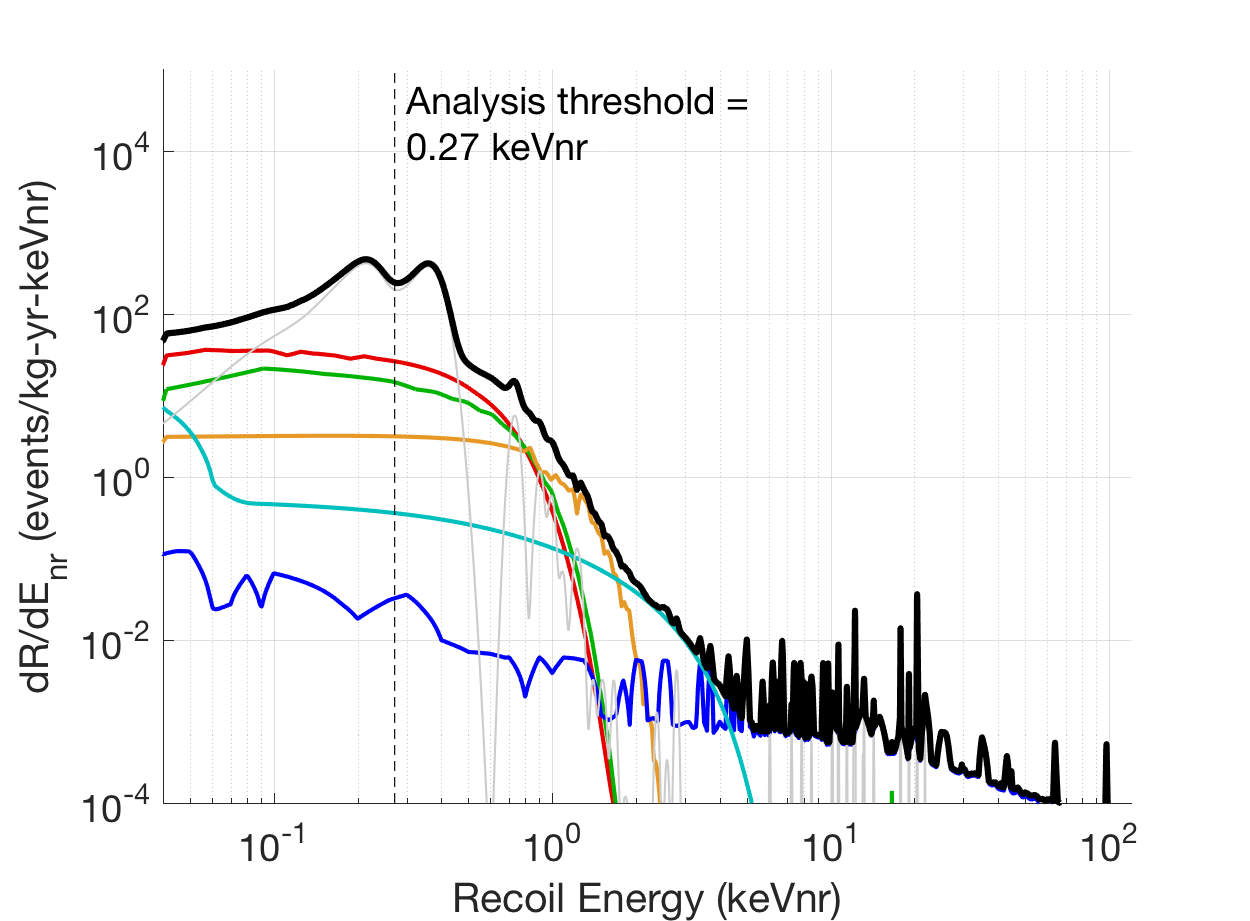}
\caption{Background spectra, before (left) and after (right) analysis cuts in Si (top) and Ge (bottom) iZIP detectors, shown as a function of nuclear recoil energy (keVnr) to allow direct comparison of the various backgrounds to the natural dark matter interaction energy scale. Thick black lines represent the total background rates. Electron recoils from Compton gamma-rays, $^3$H and $^{32}$Si are grouped together  (red). The Ge activation lines (grey) are shown convolved with a 10\,eV r.m.s.\ resolution. The remaining components are surface betas (green), surface \(^{206}\)Pb recoils (orange), neutrons (blue) and CE$\nu$NS (cyan).}
\label{fig:iZIPBackgroundSpectra}
\end{figure*}

To model the effect of applying the analysis cuts to the data, each background spectrum is multiplied with the appropriate energy-dependent cut-passing efficiency. The right-hand panels in Figs.~\ref{fig:HVBackgroundSpectra} and~\ref{fig:iZIPBackgroundSpectra} illustrate the resulting background spectra after application of the cuts. It can be seen that certain populations, \textit{e.g.},\ the \(^{206}\)Pb recoils (orange), are significantly suppressed because they are exclusively located at the surfaces. For the iZIPs (Fig.~\ref{fig:iZIPBackgroundSpectra}), the combination of fiducial-volume and yield cuts rejects enough of the surface and ER backgrounds that the coherent elastic neutrino-nucleus scattering (CE$\nu$NS) signal dominates at some energies.  

The cuts are also used to determine a signal acceptance efficiency, which is mildly energy dependent and is near 85\% for the HV detectors and 75\% for the iZIP detectors, as can be inferred by comparing the CE$\nu$NS (cyan) and neutron (blue) spectra before and after cuts.

\clearpage

\subsection{Analysis Thresholds}
The nuclear recoil energy range used in this analysis for determining the experimental sensitivity extends from a lower analysis threshold  energy up to 25(120)\,keV for HV(iZIP) detectors.  The upper limit of the energy range is based on the recoil spectra of the particle masses for which the HV(iZIP) detectors are best suited. The lower analysis threshold is determined based on a combination of two criteria: 
\begin{enumerate}
\item The phonon sensor energy resolution: \(\sigma_{Ph}\).
\item The Luke-Neganov phonon energy created by a single electron-hole pair moving across the full potential difference in the detector: $e\Delta V$.
\end{enumerate}

In the total phonon energy scale, an energy threshold of \(7\sigma_{Ph}\) is used to keep the number of noise events to a negligible value. Similarly, an energy threshold of $e\Delta V$ is used to suppress events due to non-radiogenic backgrounds, such as infrared photons incident on the detectors or dark current through the biased detectors. Such events (collectively referred to as leakage current) can mimic low-energy nuclear recoils due to the Luke-Neganov phonons produced by the motion of ionized excitations traversing the detector. If the leakage current is found to exceed the nominal value of $10^{-21}$\,A, the analysis threshold can be raised to multiples of $e\Delta V$ to suppress the contribution from multiple excitation coincidences.

The larger of the \(7\sigma_{Ph}\) and $e\Delta V$ quantities for each detector is chosen as the analysis threshold in the ``total phonon'' energy scale and is converted to its nuclear recoil equivalent using the appropriate yield function for the detector material. Table~\ref{tab:AnalysisThreshold} lists the projected analysis thresholds used for the HV and iZIP Si/Ge detectors. Once experimental data is in hand, analysis thresholds will be chosen to reflect the optimal values based on the data. Additionally, potential effects of leakage currents larger than the nominal value can also be mitigated by reducing the bias voltage, which will also be tuned based on initial data.

\begin{table}[htp]
\begin{tabular}{lcccc} \hline
& \(7\sigma_{Ph}\) & \(e\Delta V\) & \multicolumn{2}{c}{Analysis threshold}  (eV) \\
Detector &  (eV) &   (eV)& \(E_{Ph}\) & \(E_{nr}\) \\
\hline
Si HV & 35 & 100 & 100 & 78\\
Ge HV & 70 & 100 & 100 & 40\\\hline
Si iZIP & 175 & 8 & 175 & 166\\
Ge iZIP & 350 & 6 & 350 & 272\\ \hline
\end{tabular}

\caption{Summary of the energy values used in determining the analysis thresholds in the HV and iZIP Si and Ge detectors. The projected analysis threshold in the ``total phonon'' energy scale (\(E_{Ph}\)) is the maximum of the first two columns (\(7\sigma_{Ph}\) and \(e\Delta V\)). \(E_{Ph}\) is then converted into a nuclear recoil energy scale value (\(E_{nr}\)) by inverting Equation~\ref{eq:pt}.}
\label{tab:AnalysisThreshold}
\end{table}

\section{Calculating the experimental sensitivity}
\label{sec:sensitivity}
\begin{figure*}[th!]
\begin{center}
\includegraphics[width=.65\textwidth]{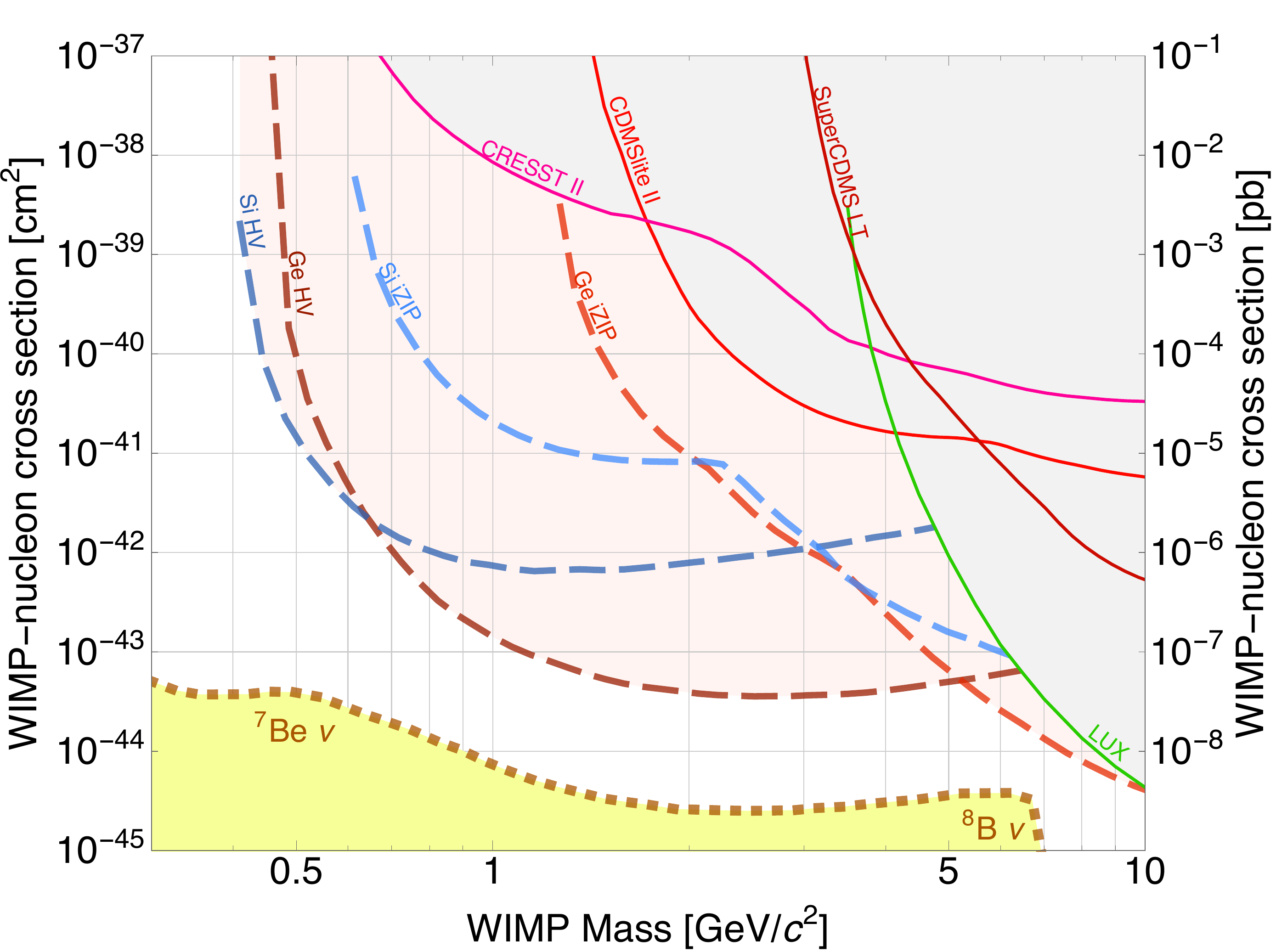} \\
\caption{Projected exclusion sensitivity for the SuperCDMS SNOLAB direct detection dark matter experiment. The vertical axis is the spin-independent WIMP-nucleon cross section under standard halo assumptions~\cite{Lewin1996}, and the horizontal axis is the WIMP mass, where WIMP is used to mean any low-mass particle dark matter candidate. The blue dashed curves represent the expected sensitivities for the Si HV and iZIP detectors and the red dashed curves the expected sensitivities of the Ge HV and iZIP detectors. These sensitivity limits are determined using the optimum interval method~\cite{Yellin:2002prd,Yellin:2007arxiv}, which does not incorporate any knowledge of the specific disposition and source of background events observed during the experimental operation. The solid lines are the current experimental exclusion limits in the low-mass region, from the CRESST-II~\cite{Angloher:2016epjc}, SuperCDMS~\cite{Agnese:2015nto,Agnese:2014PRL} and LUX~\cite{Akerib:2016prl} experiments. The dotted orange line is the dark matter discovery limit from Ref.~\cite{Ruppin:2014prd}, which represents the cross-section at which the interaction rate from dark matter particles becomes comparable to the solar neutrino coherent elastic scattering rate. }
\label{fig:OI-Limits}
\end{center}
\end{figure*}

The expected sensitivity curves for the total experimental exposure are calculated from the post-cut background spectra and analysis thresholds. First, a simulated set of events is drawn randomly from the background spectrum, with the total number of events given by the experimental exposure and the background normalization. The expected spectrum for a given dark matter particle mass is determined, assuming spin-independent interactions and the standard halo model\cite{Lewin1996,Agnese:2015nto}, and the post-cut efficiency is applied to it. An optimum interval calculation is then performed using the background events and dark matter spectrum to determine a 90\% C.L.\ exclusion cross section~\cite{Yellin:2002prd,Yellin:2007arxiv}. The process is repeated over the mass range of interest to obtain an exclusion curve based on the single background pseudo-experiment. Exclusion curves for a large number of such pseudo-experiments are calculated, and the median 90\%~C.L.\ exclusion cross section at each mass is chosen as the cross-section sensitivity. The resulting exclusion sensitivity reach in WIMP-nucleon cross-section/WIMP mass parameter space is presented in Fig~\ref{fig:OI-Limits}. 

Because the optimum interval method sets the limit based on a single energy interval, it can result in kinks in the limit curve at a particle mass at which the optimum transitions from one interval to another.  In particular, the kink in the Si iZIP sensitivity curve in Fig.~\ref{fig:OI-Limits} near a mass of 2\GeVcc is due to the transition from a background-free interval above 2\,keV to a background-limited interval.

The optimum interval method is a conservative approach for calculating a sensitivity limit since it does not implement background subtraction. With this type of analysis, the HV detectors will approach these sensitivities within two years of operation. Due to their background discrimination capabilities, the iZIP detectors can run for more than five years without reaching background limitations, thus achieving better sensitivity at higher masses.

As the design progresses, and background expectations are better understood, we will update the sensitivities based on likelihood analyses incorporating background subtraction, which allow the sensitivity to improve over the lifetime of the experiment. Background subtraction benefits from independent knowledge of the background levels determined through material assay of components and use of the complementarity of the HV and iZIP detectors' response to ER and NR events. 

Furthermore, in this paper we have assumed a continuous ionization model. In the future we will incorporate charge quantization into the sensitivity analysis, which in conjunction with the excellent detector phonon resolution should allow ER/NR discrimination in the HV detectors, leading to a significant improvement in sensitivity.

\section{Dependence of Sensitivity on Input Parameters}
\label{sec:Discussion}

In this section, we show the dependence of the experimental sensitivity on some of the less constrained assumptions. We studied variations in cosmogenic background rates and also in ionization yield modeling.  Only the HV studies are presented because the iZIP sensitivity curves were found to be largely insensitive to changes in background and ionization yield modeling for WIMP masses above $\sim$\,$2\GeVcc$. The iZIP's insensitivity to the varied inputs is due to a combination of the excellent ER/NR discrimination, that result in a sensitivity that is exposure-limited in this mass range, and the low applied bias voltages that minimize the contribution of Luke-Neganov phonons to the nuclear recoil signal.

\subsection{Parametric Background Variations}
\begin{figure*}[htp]
\begin{center}
\includegraphics[width=0.45\textwidth]{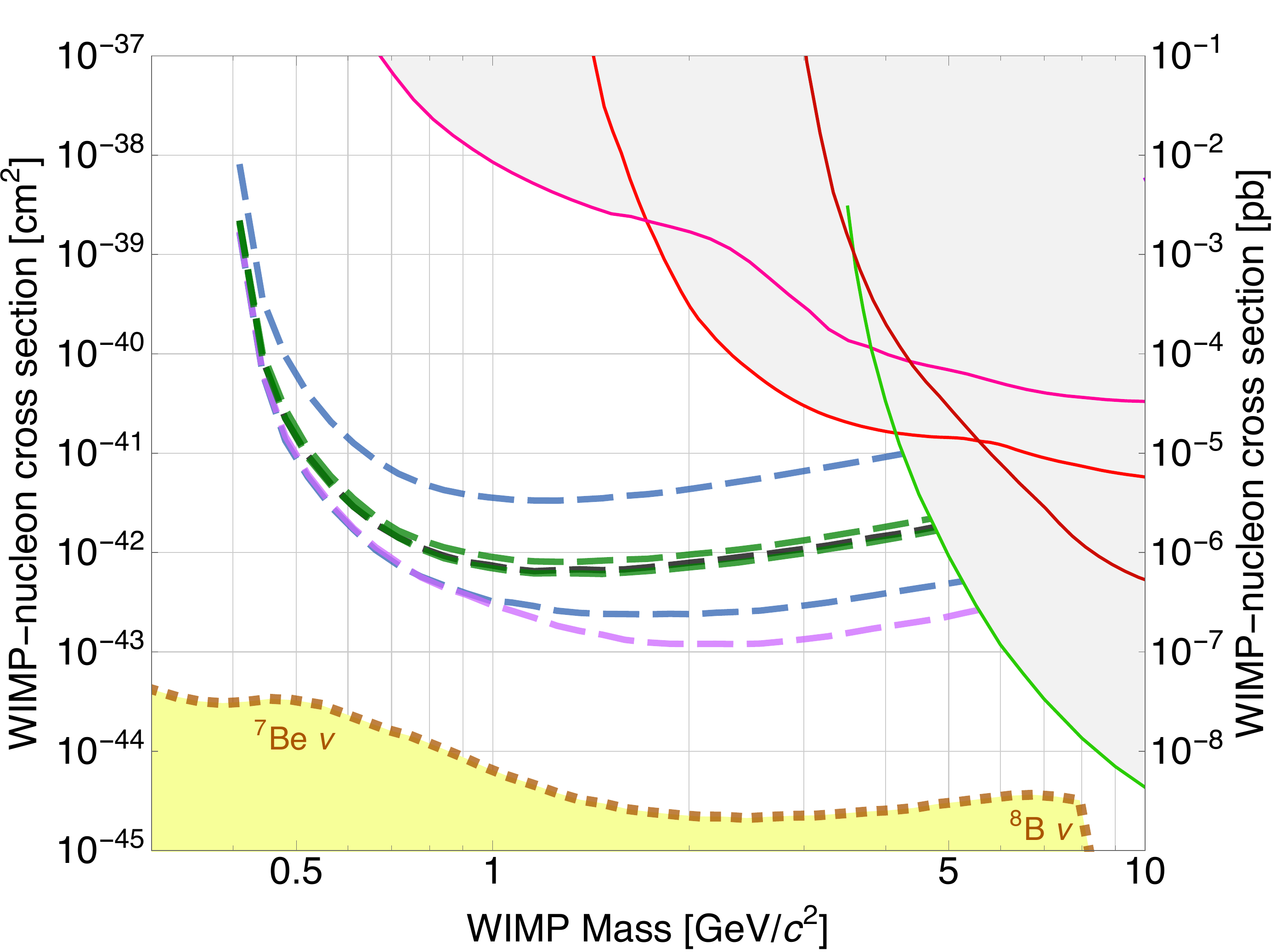}
\includegraphics[width=0.45\textwidth]{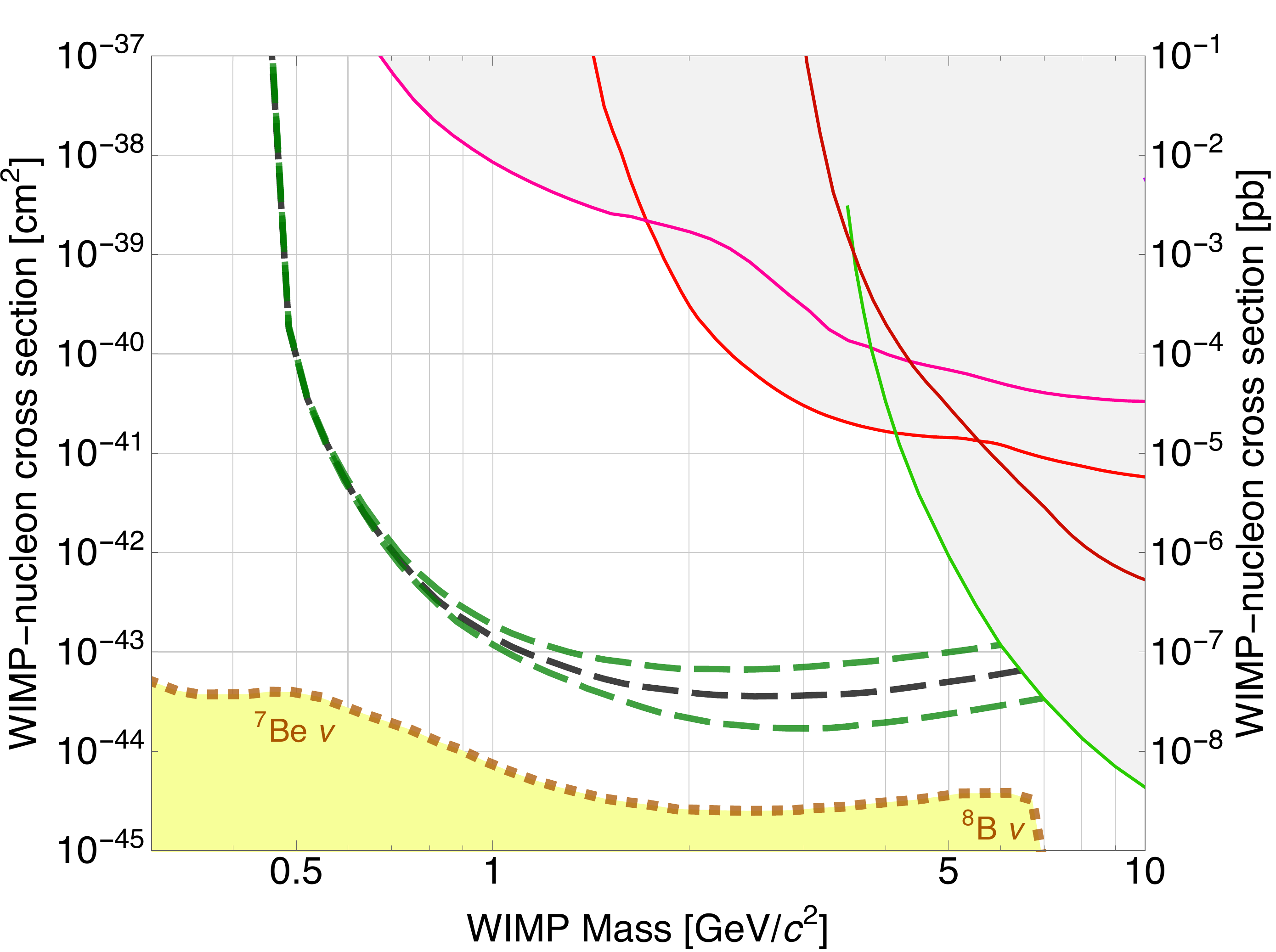}
\end{center}
\caption{Si (left) and Ge (right) HV detector sensitivities for different background assumptions. The black dashed lines are the sensitivities for the nominal assumptions (same as Fig.~\ref{fig:OI-Limits}). The green dashed lines correspond to varying the $^3$H contamination from zero (lower curve) to 3$\times$ the nominal exposure (upper curve) while keeping all other backgrounds at their nominal values. For Si, the blue dashed lines correspond to varying the $^{32}$Si contamination from zero (lower curve) to 10$\times$ the nominal value (upper curve), while keeping the $^3$H at its nominal value. The purple dashed line is the expected sensitivity if both the $^3$H and $^{32}$Si contamination levels are zero.  The solid lines are the same experimental sensitivities presented in Fig.~\ref{fig:OI-Limits}.}
\label{fig:SensitivityBkgScanHV}
\begin{center}
\includegraphics[width=0.45\textwidth]{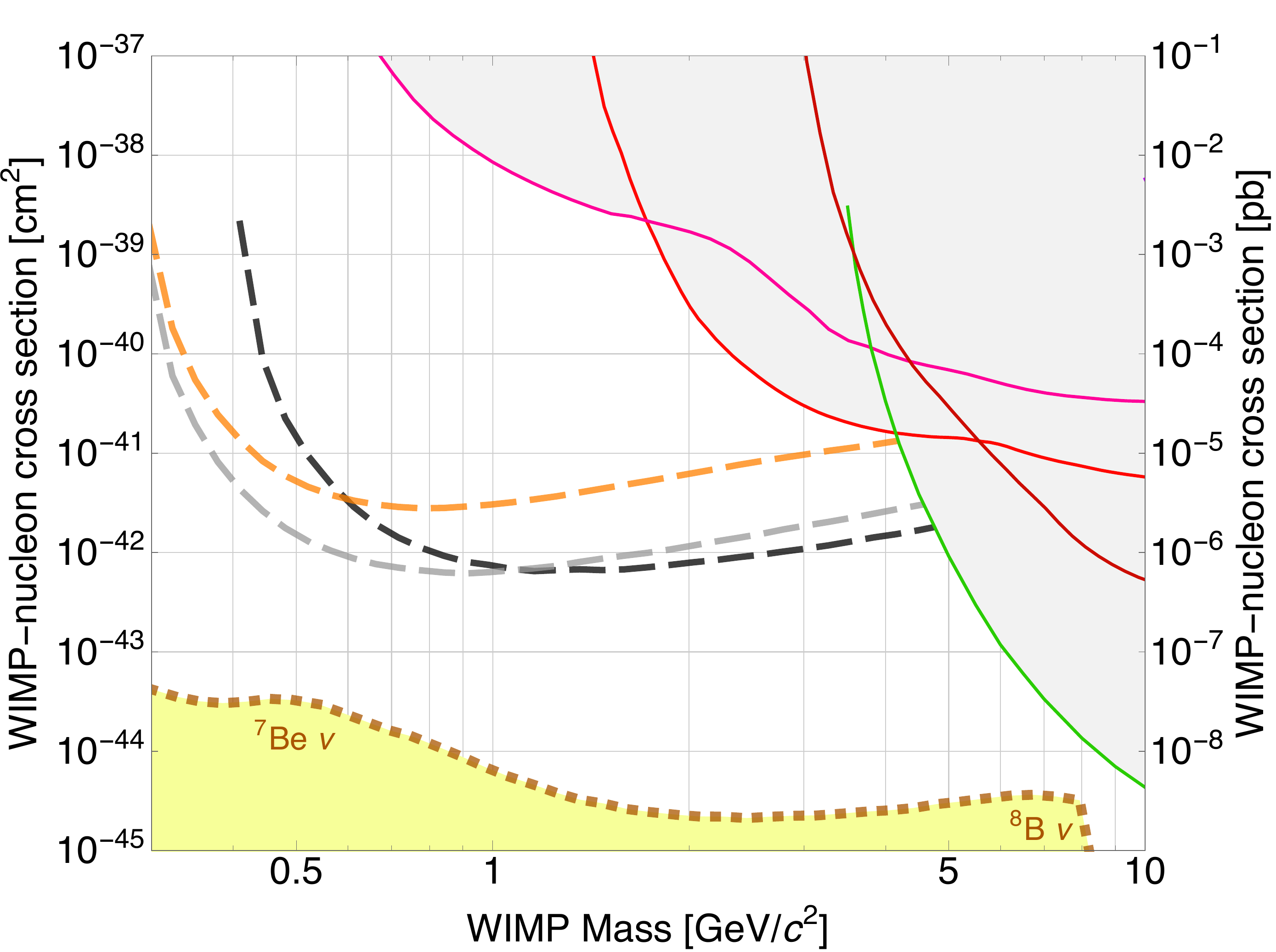}
\includegraphics[width=0.45\textwidth]{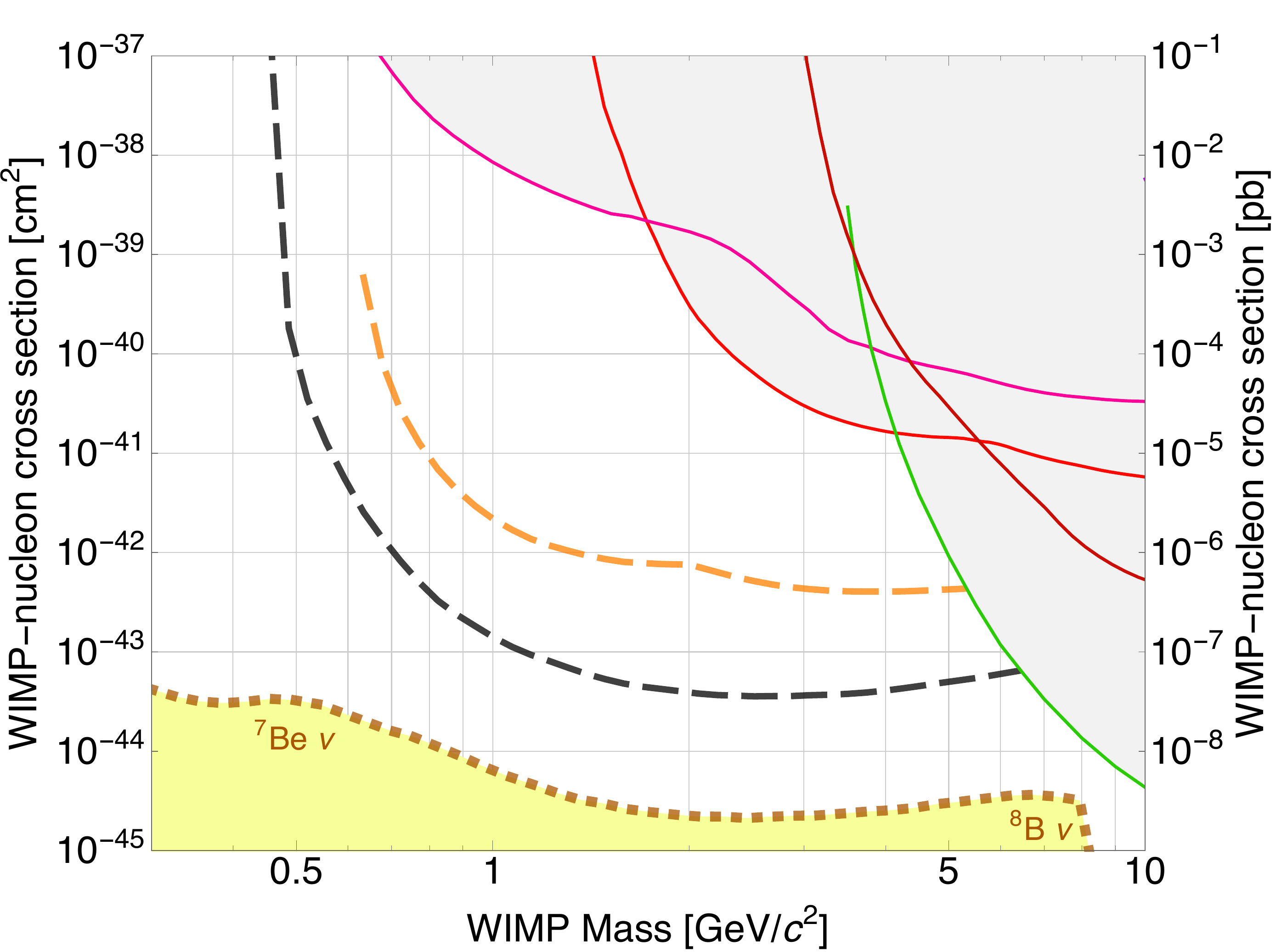}
\end{center}
\caption{Si (left) and Ge (right) HV detector sensitivity studies pertaining to yield and analysis thresholds. The black dashed lines are the sensitivities for the nominal assumptions (same as Fig.~\ref{fig:OI-Limits}). The orange dashed curves correspond to operating the detectors with zero voltage bias. For Si, the gray dashed curve shows the sensitivity using the standard Lindhard ionization yield model with an energy cutoff of \(E_{y0}=40\,\mathrm{eV}\). Solid lines are the same experimental sensitivities presented in Fig.~\ref{fig:OI-Limits}.}
\label{fig:SensitivityYVScanHV}
\end{figure*}

The effect of varying the background assumptions is shown in Fig.~\ref{fig:SensitivityBkgScanHV}, where the black curves are the nominal SuperCDMS SNOLAB sensitivities presented in Fig.~\ref{fig:OI-Limits}. For the Ge and Si HV detectors, we vary the $^3$H background by increasing the sea-level cosmogenic exposure period from the nominal value of 60 days to 180 days.  This results in an increase of $\sim$3$\times$ in the $^3$H rate. We also consider the limiting case of no $^3$H background.
These values were chosen to represent the extremes of the possible $^3$H contamination. Tritium is a dominant background for Ge HV detectors in the nominal scenario, thus the sensitivity at higher WIMP masses is affected by the increase in $^3$H background. The zero-tritium sensitivity curve is limited by the next-highest background, predominantly the Compton ERs from contamination of the material surrounding the detectors as described in Section~\ref{sec:volumesources}. The effect of varying $^3$H in Si is small because it is a sub-dominant background to $^{32}$Si.
 
For the Si HV detectors, we vary the dominant $^{32}$Si background level from a factor of ten higher than nominal to zero, to take into account potential variations in the $^{32}$Si content within the raw Si source material. We also show a scenario with no $^3$H or $^{32}$Si, which is limited by the next-highest background, predominantly the Compton ERs from contamination in the material surrounding the detectors as described in Section~\ref{sec:volumesources}.

\subsection{Ionization Yield Studies}

SuperCDMS SNOLAB detectors aim to be sensitive to nuclear recoils as low as tens of eV, energies for which no ionization yield experimental data is currently available. In calculating the experimental sensitivity we made assumptions about the functional form of the nuclear recoil ionization yield below existing data (\textit{cf.}\ Sections~\ref{sec:ySi} and \ref{sec:yGe}). The effect of the uncertainty in the ionization yield response is presented in Fig.~\ref{fig:SensitivityYVScanHV}, where the black curves are the nominal SuperCDMS SNOLAB sensitivities presented in Fig.~\ref{fig:OI-Limits}.

A limiting case of the dependence of the sensitivity on the ionization yield is obtained by operating the HV detectors at a 0\,V bias (\textit{cf.}\ orange dashed line in Fig.~\ref{fig:SensitivityYVScanHV}). In this mode, no Luke-Neganov phonons are produced by the drifting of ionized excitations and consequently the total phonon energy is independent of the details of ionization yield.
As described in Section~\ref{sec:Experiment}, the large Luke-Neganov phonon signal from ERs, due to the large voltage bias across the HV detector, serves to ``stretch'' the ER background spectrum with respect to a NR spectrum. Operating the HV detectors with zero (or small) voltage bias abandons this advantage, resulting in a minimum cross-section sensitivity that is slightly higher than the nominal curves for both Si and Ge detectors.

Because the detectors have excellent phonon resolution, they remain sensitive to the phonon-only energy of low-mass dark matter recoils. In the case of Si, the sensitivity to dark matter masses less than $\sim 0.6\GeVcc$, is enhanced. This is a consequence of how the analysis threshold is determined from the detector operation conditions.
As shown in Table~\ref{tab:AnalysisThreshold}, the analysis threshold for the nominal 100\,V bias mode is determined by the requirement to eliminate leakage current, resulting in a threshold given by \(E_{ph} = e\Delta V = 100\,\)eV. This corresponds to a nuclear recoil energy of 78\,eV. For the 0\,V bias mode, the \(e\Delta V\) term vanishes (as do any events associated with leakage current), resulting in a threshold given by \(E_{ph} = 7\sigma_{ph} = 35\,\)eV. This corresponds to a nuclear recoil energy of 35\,eV since there is no Luke-Neganov phonon production at 0\,V bias. The gained sensitivity to nuclear recoils between 35 and 78\,eV leads to the improved low-mass performance. 
Figure~\ref{fig:SensitivityYVScanHV} also presents the expected sensitivity of the Si HV detectors based on the standard Lindhard ionization yield model with an energy cutoff of \(E_{y0}=40\,\mathrm{eV}\), intended as a limiting case of a favorable ionization response, where the benefit of Luke-Neganov amplification for suppressing ER events is exploited. This calculation achieves a better sensitivity compared to both the nominal and 0\,V bias modes for Si.

For the Ge detectors, the analysis thresholds are set at 40 and 70\,eV, for the 100 and 0\,V bias modes respectively, leading to a 0\,V sensitivity curve that is worse than the nominal one for all masses. Unlike the case of the Si detectors, the analysis threshold for the Ge detectors is higher for the 0\,V bias mode as a consequence of the more favorable yield model and a worse phonon resolution than for Si.

In summary, the 0\,V and Lindhard sensitivity curves represent two extreme cases for the nuclear recoil ionization yield.  The comparison of 0\,V and 100\,V performance highlights the important roles of Luke-Neganov phonon amplification as a means to suppress ER backgrounds and to enhance sensitivity to low-energy nuclear recoils in certain cases. Most importantly, this study demonstrates that the phonon sensitivity for SuperCDMS SNOLAB HV detectors, even without the benefit of Luke-Neganov amplification, is sufficient to detect nuclear recoils from sub-GeV dark matter.

%%
%% CONCLUSION
%%

\section{Conclusion}
\label{sec:Conclusion}

In this paper, we have detailed the calculation of the sensitivity to low-mass dark matter particles for the SuperCDMS SNOLAB experiment and provided a snapshot of the initial design sensitivity~\cite{LimitPlotter}. We have also provided a set of studies that vary several key input parameters over a wide range to motivate understanding for how the sensitivity depends on the detector response and background assumptions. We have shown that there is considerable flexibility provided by the mix of detector types, and how they are operated, which will provide a robust path to detection of the lowest mass dark matter particles. The dependence of the sensitivities on backgrounds is presented conservatively, with no attempt at background subtraction. As the design progresses, and backgrounds are modeled better, we will update the sensitivities and present likelihood analyses that incorporate background subtraction, to give a more accurate projection for the initial reach for SuperCDMS SNOLAB. The ultimate goal of the experiment, to reach the solar neutrino floor, will be enabled by R\&D on improved detectors with lower energy thresholds and backgrounds.

%%
%% ACKNOWLEDGEMENTS
%%

\section{Acknowledgements}
This work is supported in part by the National Science Foundation, by the United States Department of Energy, by  NSERC Canada, and by MultiDark (Spanish MINECO). Fermilab is operated by the Fermi Research Alliance,  LLC under Contract No.\ De-AC02-07CH11359. Pacific Northwest National Laboratory is operated by Battelle for the United States Department of Energy under Contract No.\ DE-AC05-76RL01830. SLAC is operated under Contract No.\ DE-AC02-76SF00515 with the United States Department of Energy.

%%
%% BIBLIOGRAPHY
%%

\bibliography{references.bib}

\end{document}